\documentclass[iop,apj,tighten]{emulateapj}
\usepackage{apjfonts} 
\usepackage{amsmath,amstext,mathtools}
\usepackage[breaklinks,colorlinks,citecolor=blue,linkcolor=magenta]{hyperref} 
\usepackage[all]{hypcap} 
\usepackage{relsize}

\def\red#1 {\textcolor{red}{#1}\ }   
\def\blue#1 { \textcolor{blue}{\bf #1}\ }   
 
\newcommand{\dott}[1]{\skew{1.0}\dot{#1}}

\newcommand{\vel}{\upsilon}

\shorttitle{Finite and Infinite Circumbinary Disks}
\shortauthors{Mu\~noz et al.}

\begin{document}

\title{Circumbinary Accretion from Finite and Infinite Disks}
\author{Diego J. Mu\~noz$^1$, Dong Lai$^{2.3}$, Kaitlin Kratter$^3$and Ryan Miranda$^4$}
\affil{
$^1$Center for Interdisciplinary Exploration and Research in Astrophysics, Physics and Astronomy,
Northwestern University, Evanston, IL 60208, USA\\
$^2$Cornell Center for Astrophysics and Planetary Science, Department of Astronomy, Cornell University, Ithaca, NY 14853, USA\\
$^3$Tsung-Dao Lee Institute, Shanghai Jiao Tong University, Shanghai 200240, China\\
$^4$ Steward Observatory, University of Arizona, Tucson, AZ 85721, USA\\
$^5$Institute for Advanced Study, School of Natural Sciences, Einstein Drive, Princeton, NJ 08540, USA
}

\begin{abstract}
We carry out 2D viscous hydrodynamics simulations of circumbinary disk (CBD) accretion using {\footnotesize AREPO}. We resolve the accretion flow from a large-scale CBD down to the streamers and disks around individual binary components. Extending our recent studies \citep{mun19}, we consider circular binaries with various mass ratios ($0.1\leq q_{\rm{b}}\leq1$) and study accretion from ``infinite'', steady-supply disks and from finite-sized, viscously spreading tori.  For ``infinite'' disks, a global steady state can be reached, and the accretion variability has a dominant frequency ${\sim}0.2\Omega_{\rm{b}}$ for $q_{\rm{b}}>0.5$ and $\Omega_{\rm{b}}$ for $q_{\rm{b}}<0.5$,  ($\Omega_{\rm{b}}$ is the binary angular frequency). We find that the accretion ``eigenvalue'' $l_0$ -- the net angular momentum transfer from the disk to the binary per unit accreted mass -- is always positive and falls in the range ($0.65$-$0.85)a_{\rm b}^2\Omega_{\rm{b}}$ (with $a_{\rm{b}}$ the binary separation), depending weakly on the mass ratio and viscosity. This leads to binary expansion when $q_{\rm{b}}\gtrsim0.3$. Accretion from a finite torus can be separated into two phases: an initial transient phase, corresponding to the filling of the binary cavity, followed by a viscous pseudo-stationary phase, during which the torus viscously spreads and accretes onto the binary. In the viscous phase, the net torque on the binary per unit accreted mass is close to $l_0$, the value derived for ``infinite'' disks. We conclude that similar-mass binaries accreting from CBDs gain angular momentum and expand over long time scales. This result significantly impacts the coalescence of supermassive binary black holes and newly formed binary stars. We offer a word of caution against conclusions drawn from simulations of transient accretion onto empty circumbinary cavities.
\end{abstract}

\maketitle

\section{Introduction}
Circumbinary disk (CBD) accretion plays an important role in the evolution
of many types of binary systems, ranging from young binary stars to
massive binary black holes (MBBH). In these systems, the combined
effects of accretion and binary-disk gravitational interaction dictate
the long-term evolution of the binary orbit. Numerical simulations are
required to understand the accretion process, since the flow is
complex and covers a wide range of scales (from the outer CBD,
through accretion streams, to circum-single disks onto individual
binary components). Moreover, to determine the secular effect of 
accretion on the orbital evolution of the binary, long-term simulations are
needed in order to average out the rapid flow variability and
transient features. For these reasons, numerical simulations of
circumbinary accretion are challenging, and only recently have consistent
results on the long-term evolution of accreting binaries 
begun to emerge (see \citealp{mun19} and \citealp{moo19}).

The most important byproduct
of binary-disk interaction is the change in the binary's semi-major axis $a_{\rm b}$.
The early theoretical and computational works of \citet{art94} and \citet{art96}
concluded that binaries surrounded by gas disks  
evolve toward coalescence. These works, however, ignored the effect of accretion, assuming
that the cavity carved by the tidal potential was empty enough to partially or totally
suppress accretion onto the central objects. Subsequent work expanded upon
the original findings of Artymowicz and Lubow, always concluding that
the binary migrates inwards \citep[e.g.,][]{mac08,far14}.

Cosmological hierarchical structure formation predicts the formation
of massive binary black holes (MBBHs)
\citep[e.g.,][]{beg80,vol03}, but observations have not been able to
discern whether these binaries merge, or ``stall'' at some finite
separation.  The stalling of MBBHs has been dubbed the ``final parsec
problem'' \citep{mil03a,mil03b}, and occurs when all the dynamical
mechanisms that extract angular momentum have been exhausted. One
potential solution to this ``problem'' is the incorporation of
dissipative gas dynamical processes, such as the interaction with a
circumbinary gas disk \citep{hai09,hay09,roe12}, giving rise to an increased
fraction of gas-assisted MBBH mergers \citep[e.g.,][]{koc11,kel17a}.
Such mergers would generate gravitational waves (GWs) in the low-frequency
band \citep{hae94,wyi03}, and thus, understanding the coupling of
accreting binaries with surrounding gas disks is essential for making
meaningful predictions for GW background and event rates.

Recently, \citet{mir17} carried out 2D viscous hydrodynamical
simulations of circumbinary accretion using the {\footnotesize PLUTO} code \citep{mig07};
through a careful analysis of the angular momentum
balance in the CBD (keeping track of the viscous, advective and gravitational torques,
 they showed that the central binary gains angular momentum from
the gas. Since the  Miranda et al. simulations did not capture the entirety of the gas dynamics
inside the binary cavity (a circular region containing the binary was excised from the computational domain), 
their results should be considered as tentative. 

In \citet{mun19}, the problem was examined using the
moving-mesh code {\footnotesize AREPO} \citep{spr10a,pak16,wei19}; the
simulations fully captured the flow inside the cavity and the
circum-single disks (CSDs), resolving the flow down to separations of
$0.02a_{\rm b}$ from the individual binary components \citep[see also][]{mun16b}.
While the angular momentum transfer rate of CBDs around
circular binaries was consistent with \citet{mir17}, that of eccentric
binaries exhibited significant discrepancies, highlighting the
limitations of simulations that exclude the binary cavity.
Most importantly, \citet{mun19} showed that the (time-averaged)
angular momentum current through the
CBD $\langle\dott{ J}_{\rm d}\rangle$  is in agreement with the total torque acting 
directly on 
the binary $\langle\dott{ J}_{\rm b}\rangle$ (including both gravitational and accretion torques),
 confirming that their simulations are
in (quasi-) steady state and that accreting binaries gain angular
momentum from the disk. These results have been subsequently confirmed (and
extended to 3D inclined disks) in an independent study by
\citet{moo19} using {\footnotesize ATHENA++} \citep{sto08,whi16}.

Our previous works \citep{mun16b,mir17,mun19}
 focused on equal-mass binaries and accreting from ``infinite''
disks, where we imposed an outer boundary condition at $R_{\rm out}\gg
a_{\rm b}$ that supplied gas a constant rate $\dott{M}_0$. We have shown
that such a disk can reach a quasi-steady state, in which the
time-averaged mass and angular momentum transfer rates across the CBD
are constant.
It is natural to ask what happens to a binary that accretes
from a finite disk/torus.
In \citet{mun19}, we hypothesized that the secular angular momentum transfer rate
$\langle\dott  J_{\rm b}\rangle$ should still follow 
the net mass accretion rate $\langle \dott  M_{\rm b}\rangle$, provided that the latter 
changes slowly in time.

This work is organized as follows. In Section~\ref{sec:infinite}, we present simulations 
of accretion onto circular binaries of different mass ratios when accreting from ``infinite'' disks.
In Section~\ref{sec:finite}, we present analogous simulations of binaries supplied by finite disks or
``tori''. 
In  Section~\ref{sec:viscosity}, we examine the dependence of these results on the assumed disk
viscosity. Finally, in Section~\ref{sec:summary}, we discuss the implications of our work and 
summarize our key results.

\section{Accretion From ``Infinite'' Disks}~\label{sec:infinite}
The motivation behind simulating ``infinite'' disks
 (i.e., those with steady mass supply rate $\dott{M}_0$ at the outer boundary)
is the search for a (quasi-) steady state \citep{mun16b,mir17,mun19,dem19}.
The existence of a quasi-steady state allows for the effective erasure of the (arbitrary) initial conditions,
providing the means to truly explore the secular behavior of  accreting binaries.
Once such a steady state is reached, 
the mass supply rate $\dott{M}_0$ becomes a free scaling parameter.
In practice, evolving a system until steady state is reached can take several viscous times. Consequently,
it is useful to choose an initial condition that resembles a steady-state configuration as closely as possible.

\begin{figure}
\centering
\includegraphics[width=0.49\textwidth]{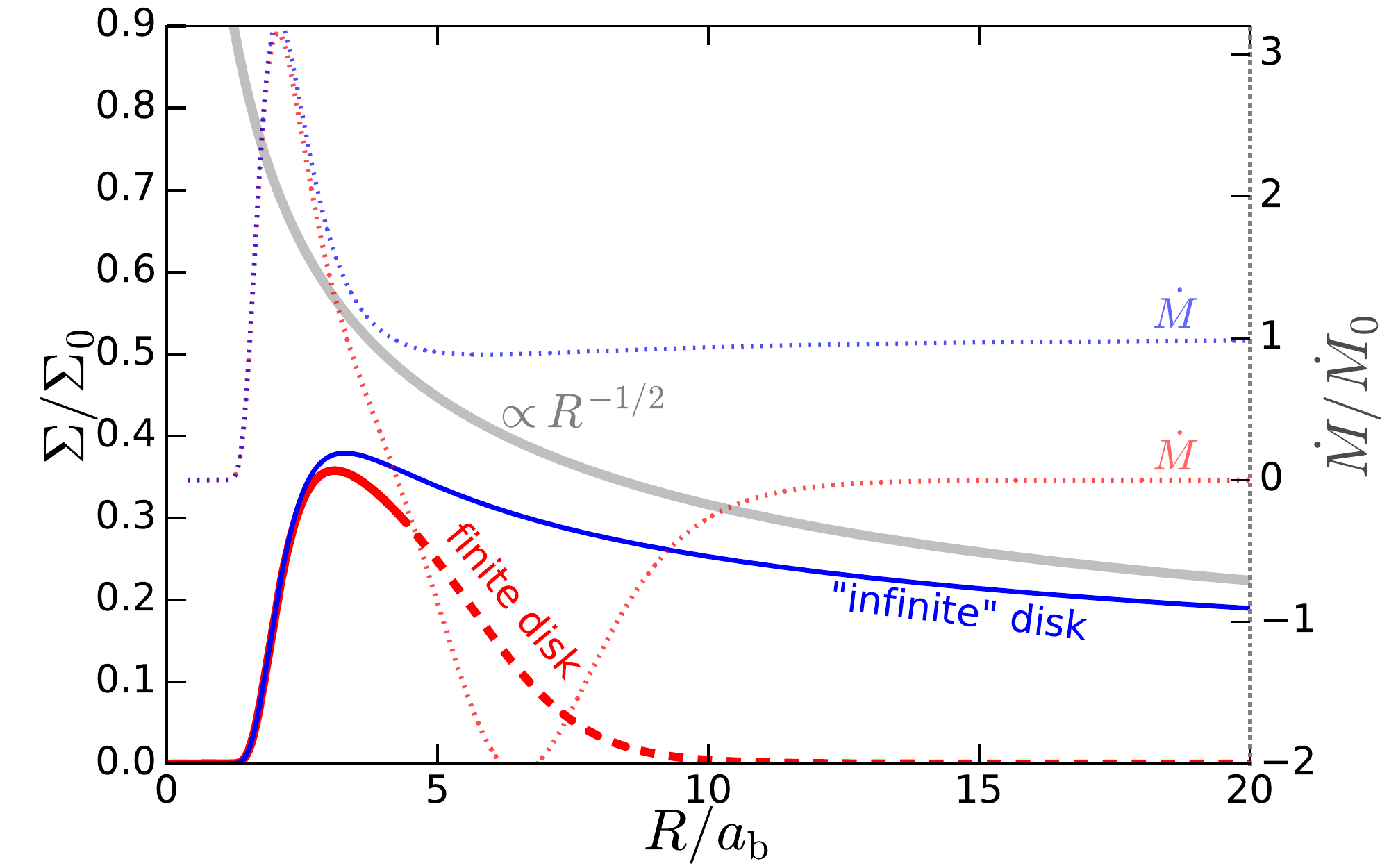}
\caption{The initial surface density profiles 
a finite disk ("torus", red) and an ``infinite'' disk (blue).
Both models share a sharp truncation at the inner edge. While the infinite disk model (Equation~\ref{eq:initial_profile_infinite})
attains a dependence on radius of the form $\Sigma\propto \nu^{-1}\propto R^{-1/2}$ at large distances,
the finite disk is tapered exponentially (Equation~\ref{eq:initial_profile_finite}) with a characteristic size
of $R_{\rm disk}=6a_{\rm b}$. The dashed region of the red curve indicates the portion of the disk
that spreads outward due to viscous stresses. Thin dotted lines (blue for infinite disk and red for finite
torus) depict the viscous accretion rate $\dott{M}(R)$ derived from $\Sigma(R)$ and $\nu\propto R^{1/2}$
and ignoring gravitational torques.
\label{fig:initial_density_profiles}}
\end{figure}

\begin{figure*}
\centering
\includegraphics[width=0.85\textwidth]{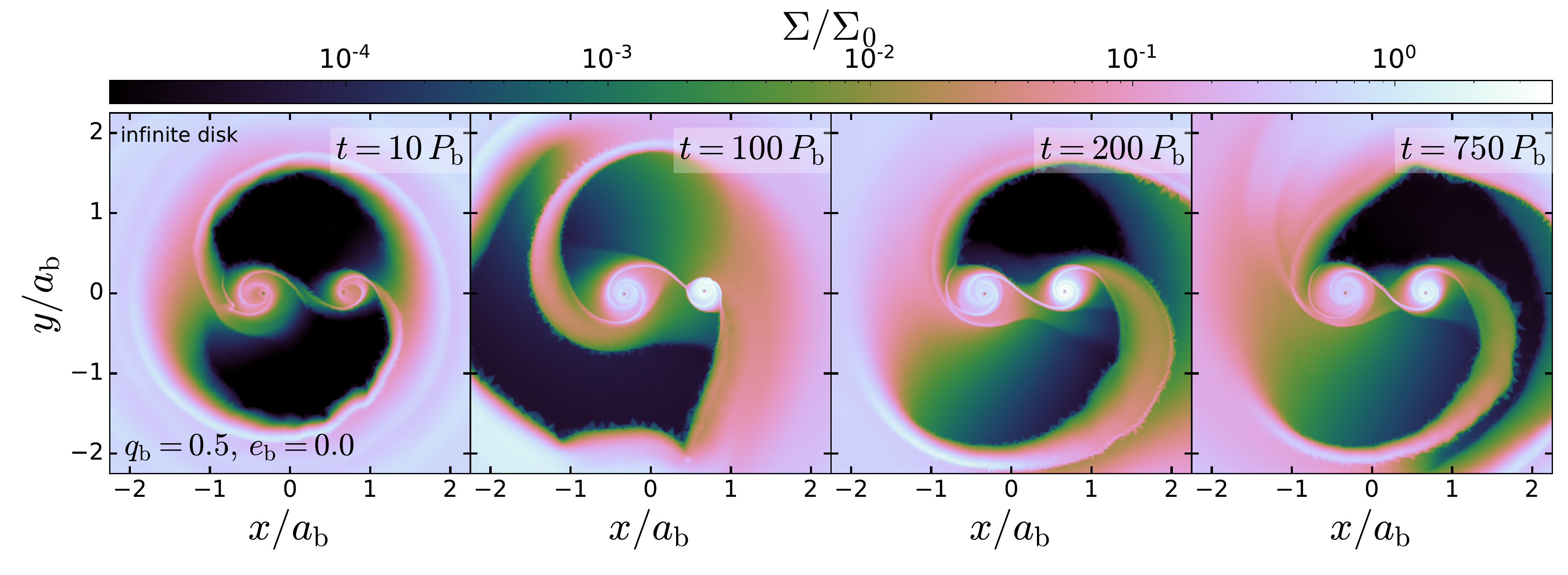}
\includegraphics[width=0.85\textwidth]{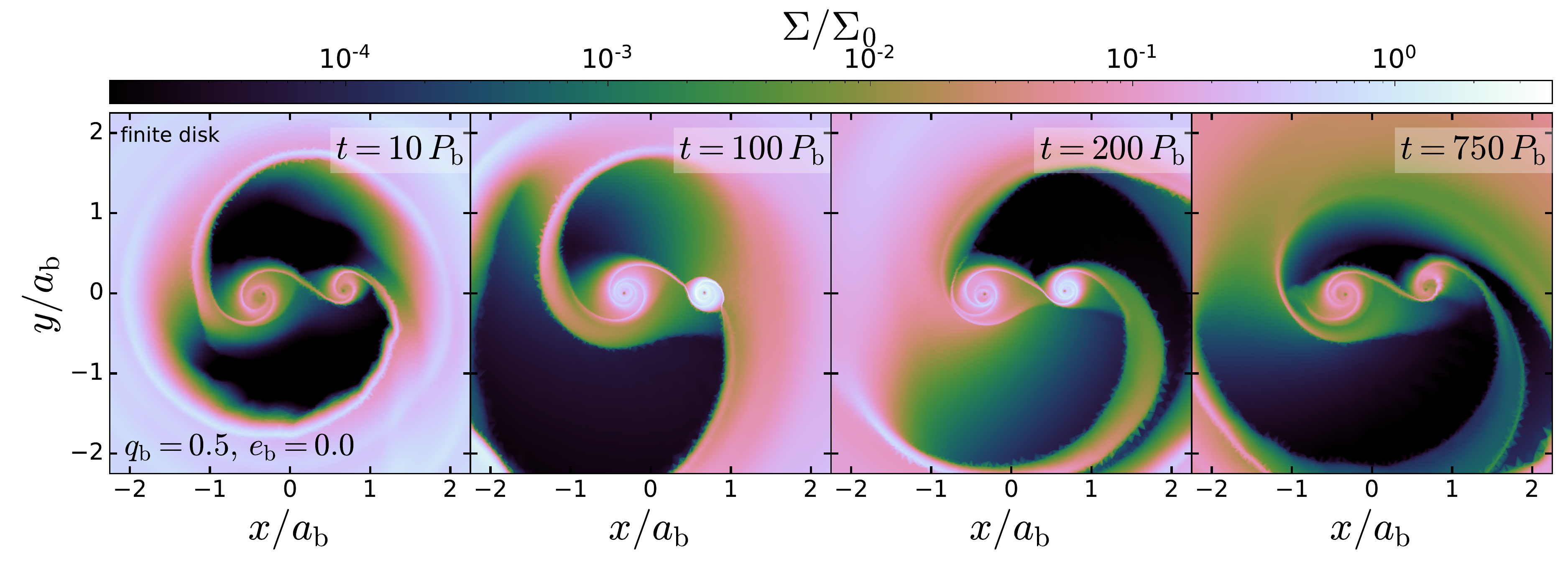}
\caption{
Accretion from an infinite disk (upper panels)  and finite torus (lower panels) 
onto a $q_{\rm b}$=0.5, $e_{\rm b}$=0 binary.  { In each case, surface density fields at different times are shown
when the $y$ coordinates of the primary (on the left) and secondary (on the right) are near zero.}
During the first few tens of orbits, the cavity (initially empty, see Fig.~\ref{fig:initial_density_profiles}) is filled 
in a similar fashion for infinite and finite disks. This is the initial transient phase. After a few hundreds of orbits,
the infinite-disk simulation approaches steady-state (the viscous time at the cavity edge is $\approx400P_{\rm b}$) while
the finite-disk simulation starts to run out of mass. This viscous-evolution phase marks a divergence between the top and bottom panels.
Note, however, that despite the lower overall density, the finite-disk case exhibits the same gas morphology as the infinite-disk simulation.
\label{fig:image_comparisons}}
\end{figure*}

\subsection{Setup and Initial Conditions}
Our setup for CBDs is similar to \citet[][Section~2]{mun19}. 
The binary is of total mass $M_{\rm b}=M_1+M_2$ and mass
ratio $q_{\rm b}=M_2/M_1$.
We use an $\alpha$
prescription for the kinematic viscosity 
$\nu$, and locally isothermal equation of state such that the disk aspect 
ratio
 $h_0=H/R$ is a constant. 
The initial disk model is described by the density profile
\begin{equation}\label{eq:initial_profile_infinite}
\Sigma(R)=f_{\rm cav}(R)\,\Sigma_0\left(\frac{R}{a_{\rm b}}\right)^{\!-\tfrac{1}{2}}\left[1-0.7\sqrt{\frac{a_{\rm b}}{R}}\right]~,
\end{equation}
where $f_{\rm cav}(R)$ is a rapidly rising function in $R$, which mimics a cavity of size $R_{\rm cav}\sim {\cal O}(a_{\rm b})$.
The precise shape of $f_{\rm cav}(R)$  is  unimportant, although it is desirable
that $f_{\rm cav}(R)\rightarrow1$ when  $R\gtrsim 5a_{\rm b}$, to guarantee that the CBD is indeed in viscous steady-state throughout most of its radial extent. 
The profile~(\ref{eq:initial_profile_infinite}) is depicted by the
blue curve in Fig.~\ref{fig:initial_density_profiles}.
The viscous accretion rate associated to this profile (blue dotted line in the figure) satisfies
$\dott{M}(R)\approx \dott{M}_0$ for $R\gtrsim 5 a_{\rm b}$, indicating that the outer regions of the CBD are in approximate steady-state from  the beginning.

The binary
affects the initial conditions of the CBD via the usual correction to the azimuthal velocity profile:
\begin{equation}\label{eq:vel_profile}
\vel_\phi^2(R)=\Omega_{\rm b}^2a_{\rm b}^2\left(\frac{a_{\rm b}}{R}\right)
\left[1 + 3\frac{Q}{R^2}\right]-c_s^2(R)\left[1 - \frac{R}{\Sigma}\frac{d\Sigma}{dR}\right]~~,
\end{equation}
where $\Omega_{\rm b}=({{\cal G}M_{\rm b}/a_{\rm b}^3})^{1/2}$ is the mean motion of the binary and
$Q\equiv \tfrac{1}{4}a_{\rm b}^2 q_{\rm b}(1+q_{\rm b})^{-2}$
is its  quadrupolar moment.  In this work, we focused on circular binaries  and explored different values of the mass ratio
$q_{\rm b}$.  {To limit the scope of this work, we explore mass ratios above $q_{\rm b}=0.1$.} 

For the disk properties, we fix the vertical aspect ratio $h_0=0.1$ and choose a fiducial disk viscosity $\alpha=0.1$ 
(the dependence of results on disk viscosity is discussed in Section~\ref{sec:viscosity}). The density scaling $\Sigma_0$ in Equation~(\ref{eq:initial_profile_infinite})
is determined by such choice of parameters:
\begin{equation}\label{eq:density_scaling}
\Sigma_0 \equiv \frac{\dott{M}_0}{3\pi \alpha h_0^2\Omega_{\rm b}a_{\rm b}^2}~~,
\end{equation}
where, in internal code units, $a_{\rm b}=\Omega_{\rm b}=\dott{M}_0=1$.

\subsubsection{Numerical Methods}\label{sec:methods}
As in \citet{mun16b} and \citet{mun19}, we carry our hydrodynamical simulations using the moving-mesh code
{\footnotesize AREPO} \citep{spr10a,pak16} in its Navier-Stokes version \citep{mun13a,mun14}. 
As an initial condition,  the resolution elements (mesh-generating points) are placed in a quasi-polar distribution in a nested fashion:
from $R=a_{\rm b}$ to $R=45 a_{\rm b}$, $N_R=475$ points are placed logarithmically in radius and $N_{\phi}=720$ points along the azimuthal direction;
from $R=45a_{\rm b}$ to $R=95 a_{\rm b}$, $N_R=62$ and $N_\phi=480$. 
Consecutive annuli of cells are interleaved, resulting in an approximately centroidal Voronoi mesh \citep{mun14}.
We impose an inflow boundary (constant accretion rate)
 at $R_{\rm out}=95a_{\rm b}$.
 In the vicinity of the
binary, the resolution is smoothly switched over to a mass-based criterion, 
with the targeted mass resolution of  
 $m_{\rm target}=5.3\times10^{-7}\Sigma_0a_{\rm b}^2$. Accretion onto the individual binary components
 is carried out within a sink region of outer edge $r_{\rm acc}=0.03a_{\rm b}$, 
 taken to be equal to the gravitational softening length of each Keplerian potential.

\begin{figure}
\centering
\includegraphics[width=0.48\textwidth]{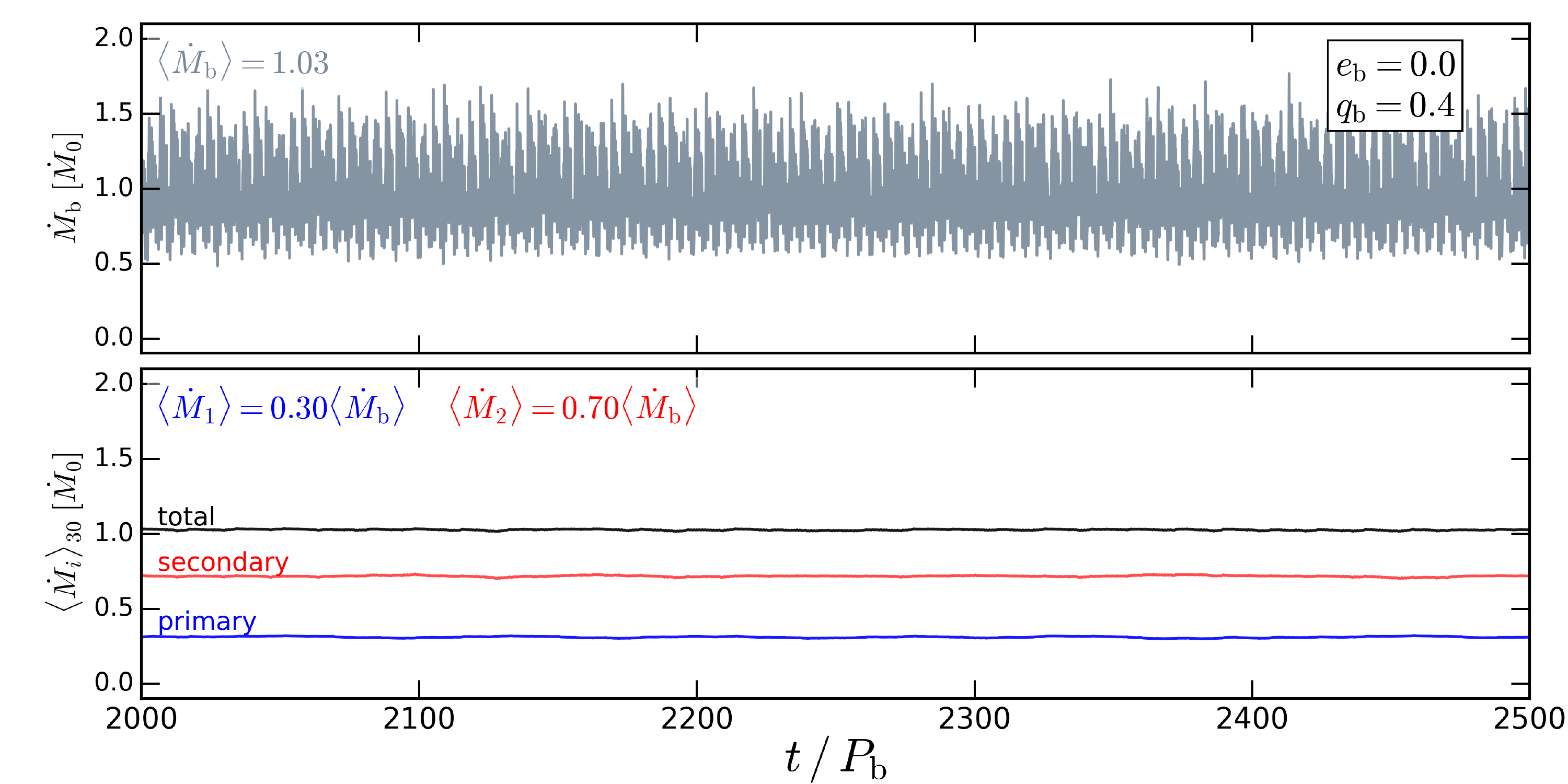}
\caption{
Accretion from an infinite disk onto a circular binary with $q_{\rm b}=0.4$ after 2000 binary orbits.
The top panel shows
the total binary accretion rate $\dott{M}_{\rm b}$, and its time-averaged value 
$\langle\dott{M}_{\rm b}\rangle\approx \dott{M}_0$. The  bottom panel shows
the running average time series (see text) of the total accretion rate
  $\langle\dott{M}_{\rm b}\rangle_{30}$ (black), the primary accretion rate 
    (blue) and the accretion rates onto the primary $\langle\dott{M}_1\rangle_{30}$ (blue)
and the   secondary $\langle\dott{M}_2\rangle_{30}$ (red).
\label{fig:steady_accretion}}
\end{figure}

\begin{figure*}
\centering
\includegraphics[width=0.44\textwidth]{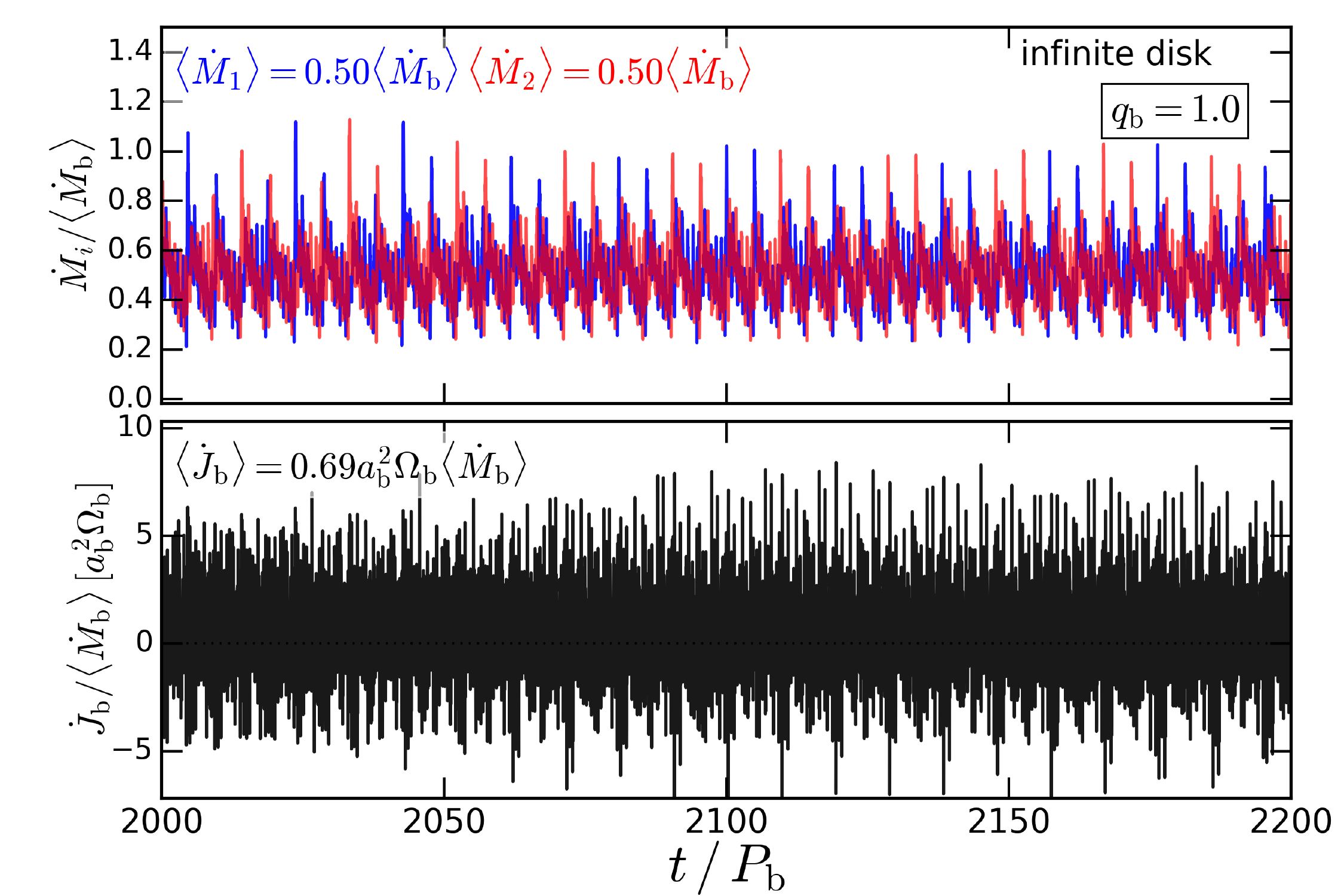}
\includegraphics[width=0.44\textwidth]{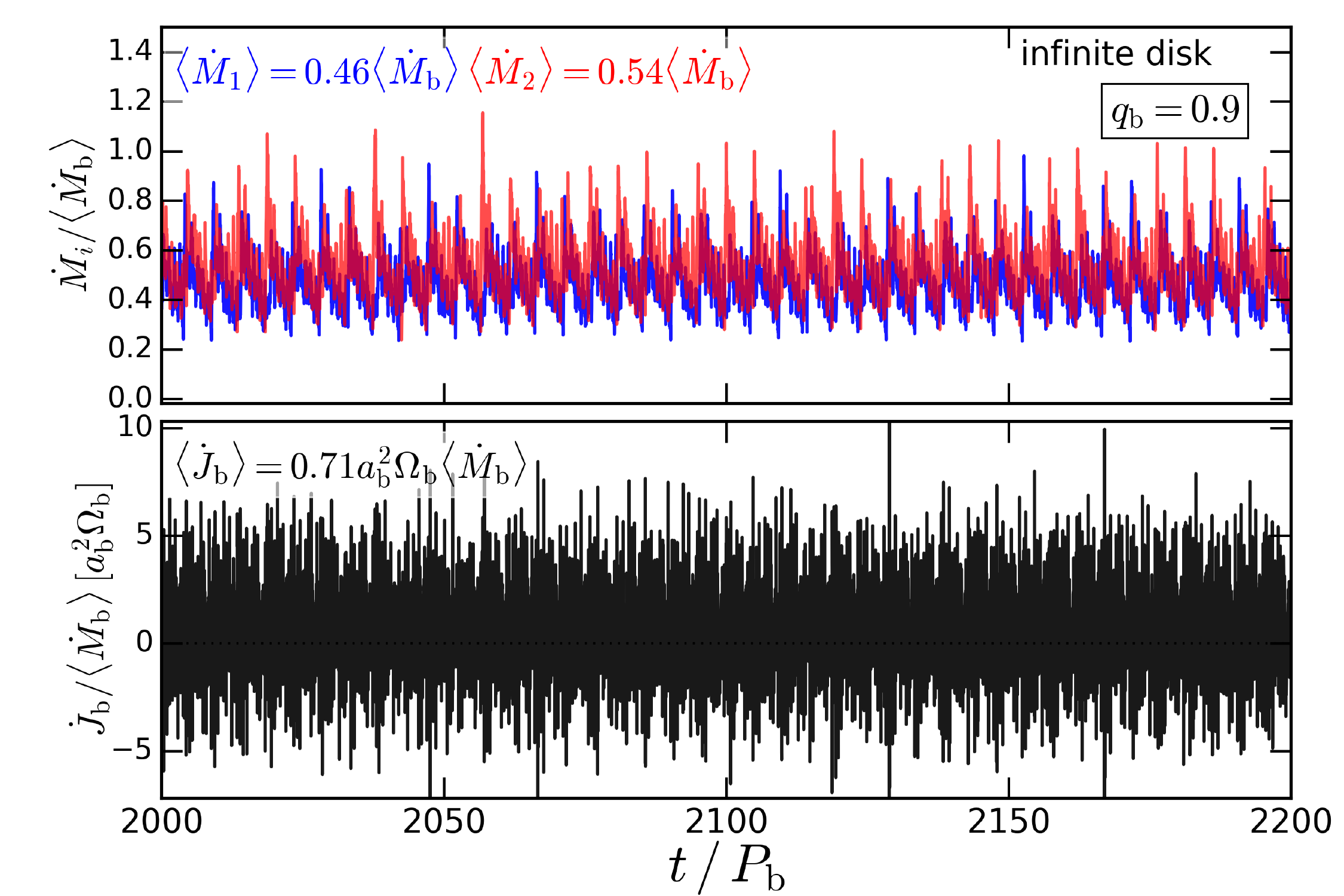}
\includegraphics[width=0.44\textwidth]{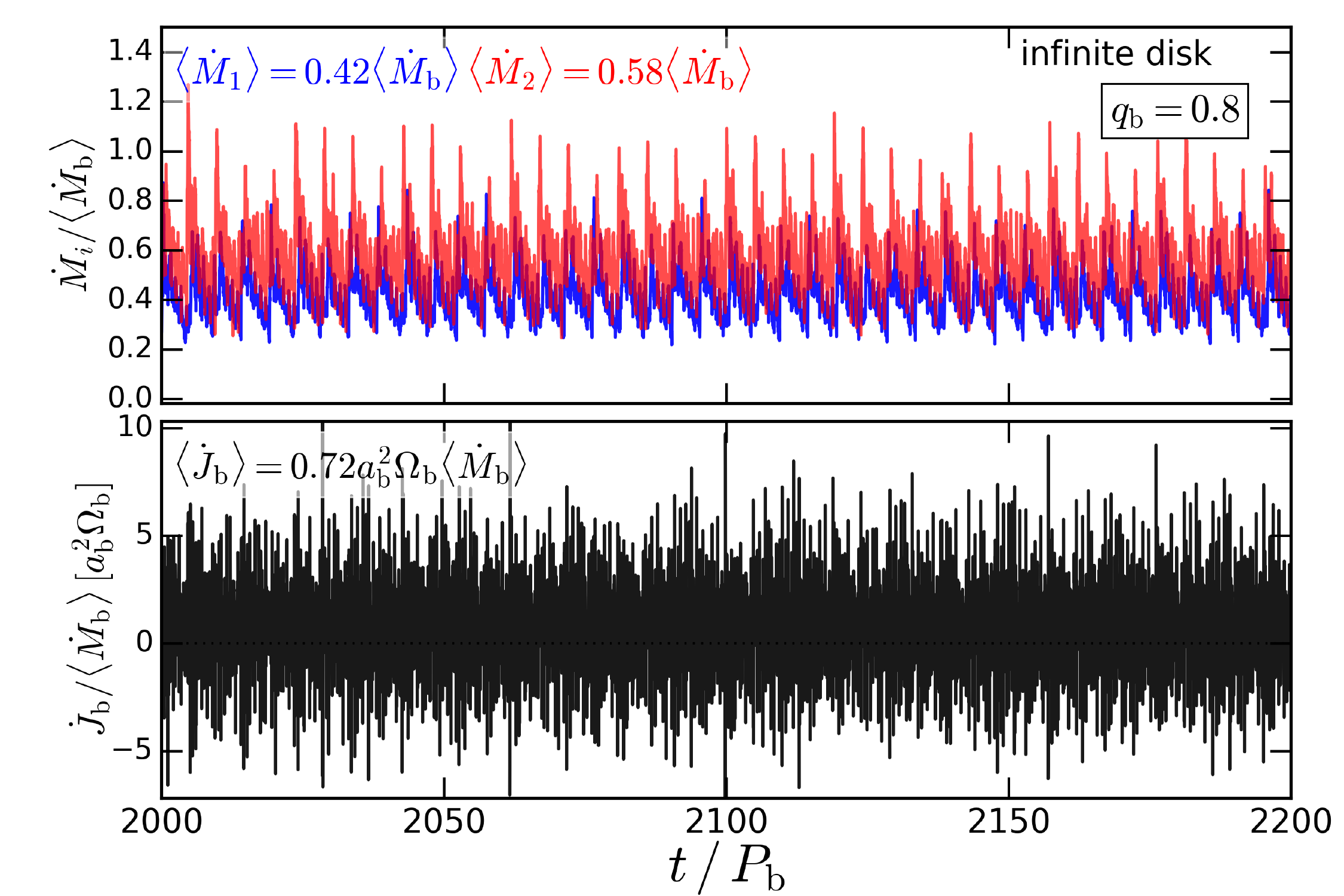}
\includegraphics[width=0.44\textwidth]{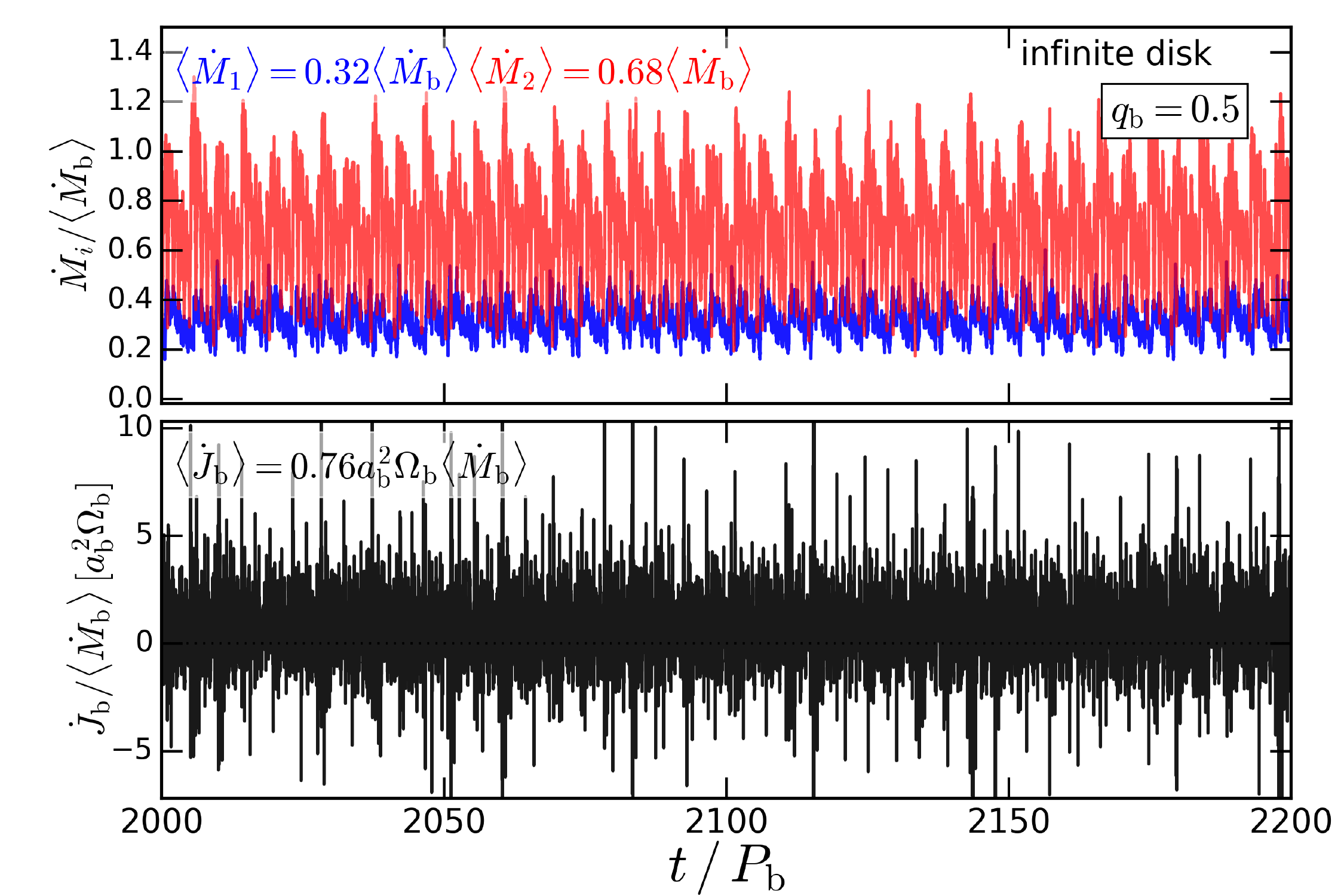}
\caption{
Accretion from ``infinite'' disks onto circular binaries of different mass ratios: $q_{\rm b}=1.0$, 0.9, 0.8 and 0.5. 
In each of the four frames, the top panels show
the stationary accretion rates onto the primary $\dott{M}_1$ (blue curves) and secondary $\dott{M}_2$ (red curves);
bottom panels show the corresponding angular momentum transfer rate $\dott{J}_{\rm b}$. From these figures, one can obtain
the time-averaged accretion ratio ${\langle\dott{M}_2\rangle}/{\langle\dott{M}_1\rangle}\geq1$ and
 the accretion eigenvalue 
$l_0\equiv {\langle\dott{J}_{\rm b}\rangle}/\dott{M}_0$.
\label{fig:accretion_rate_comparisons}}
\end{figure*}

\begin{figure*}
\centering
\includegraphics[width=0.44\textwidth]{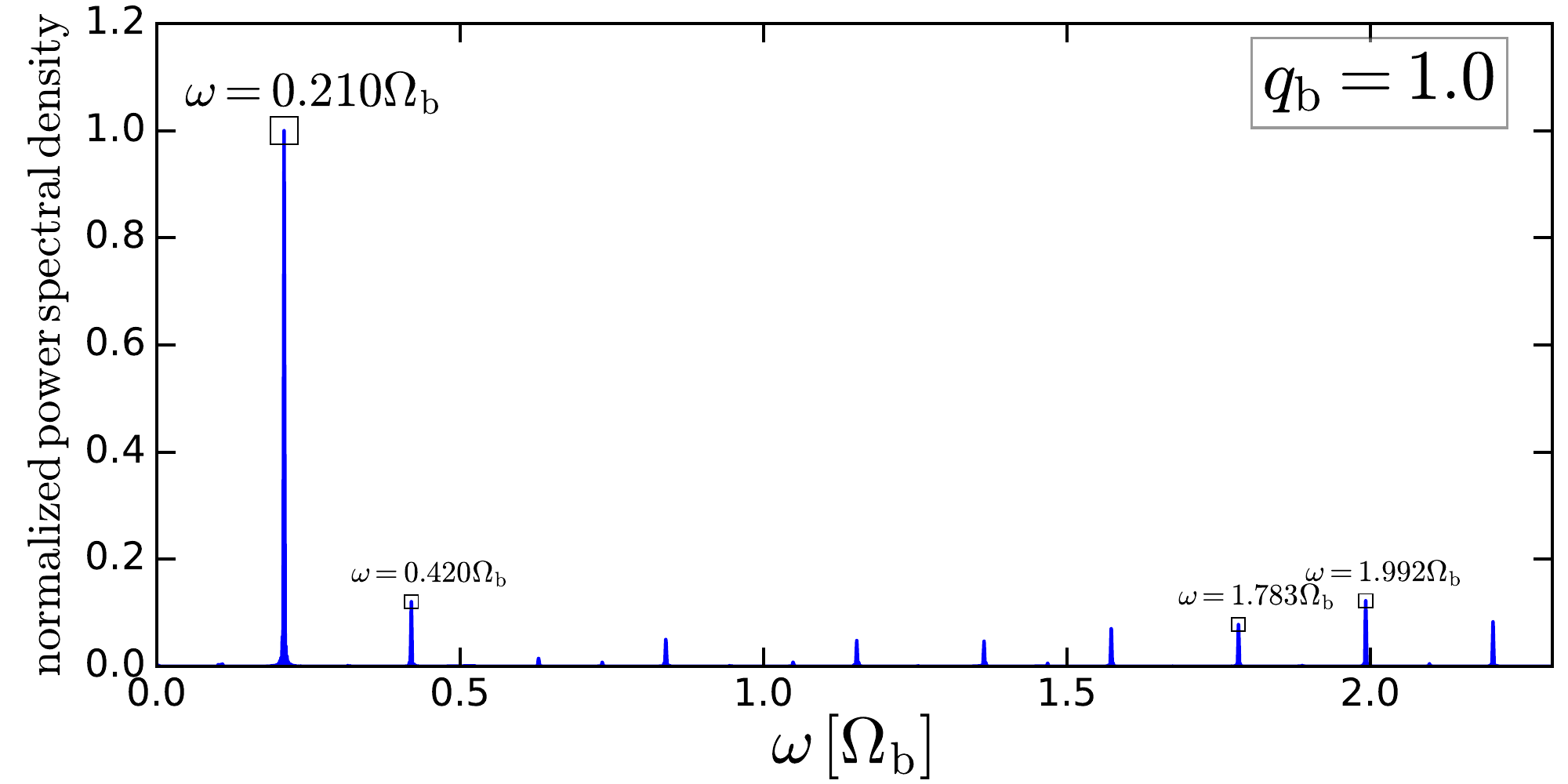}
\includegraphics[width=0.44\textwidth]{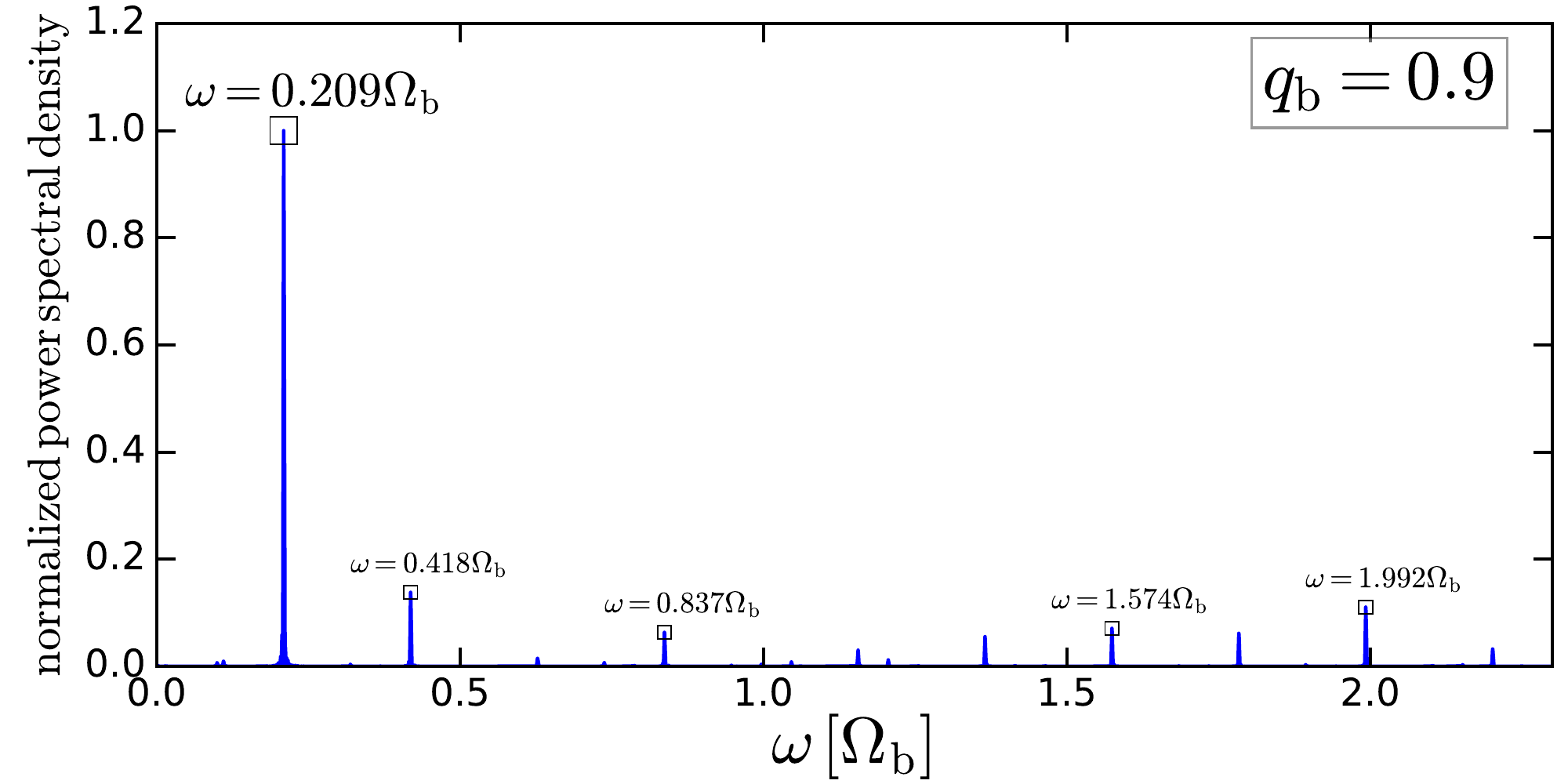}
\includegraphics[width=0.44\textwidth]{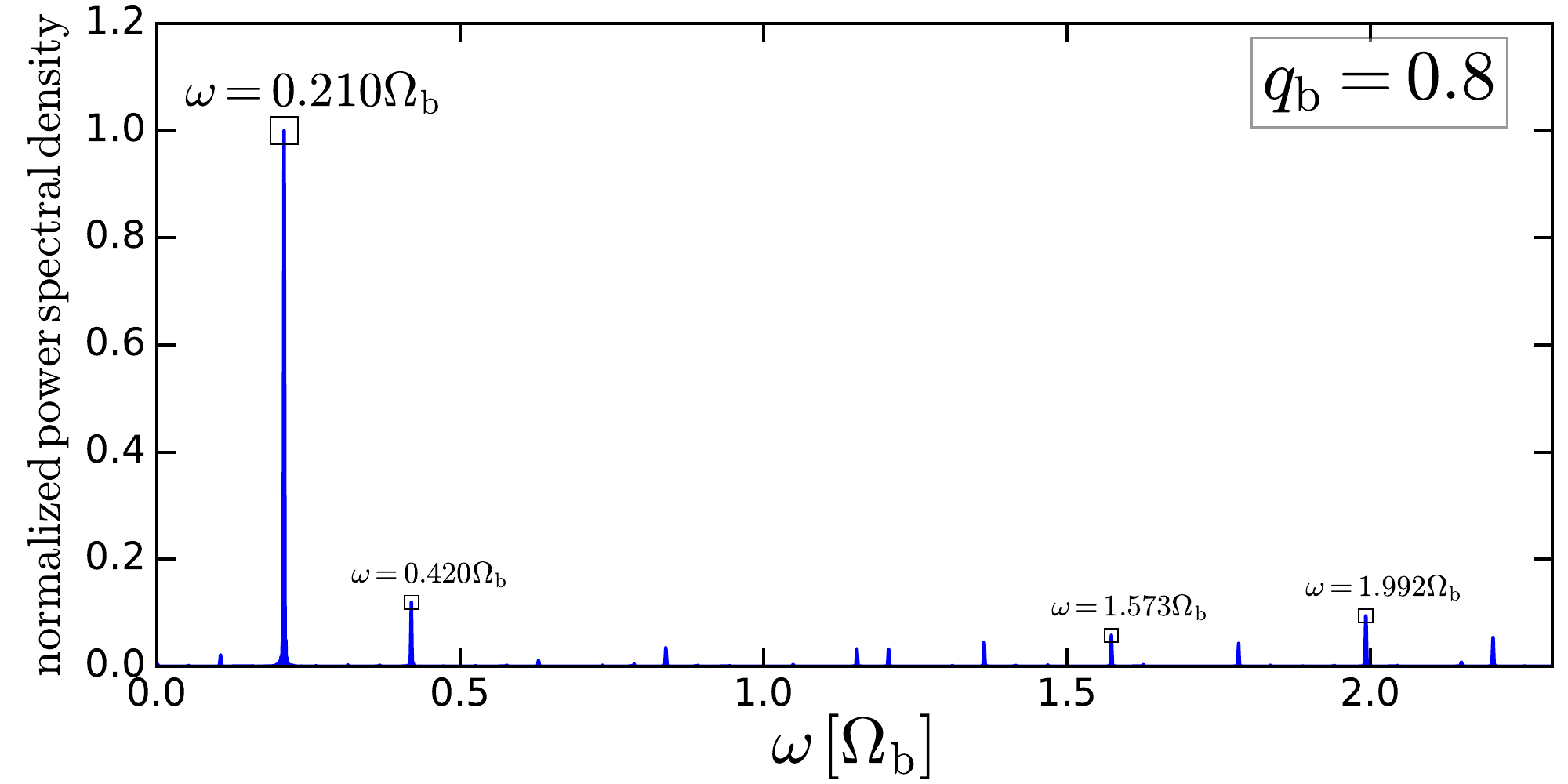}
\includegraphics[width=0.44\textwidth]{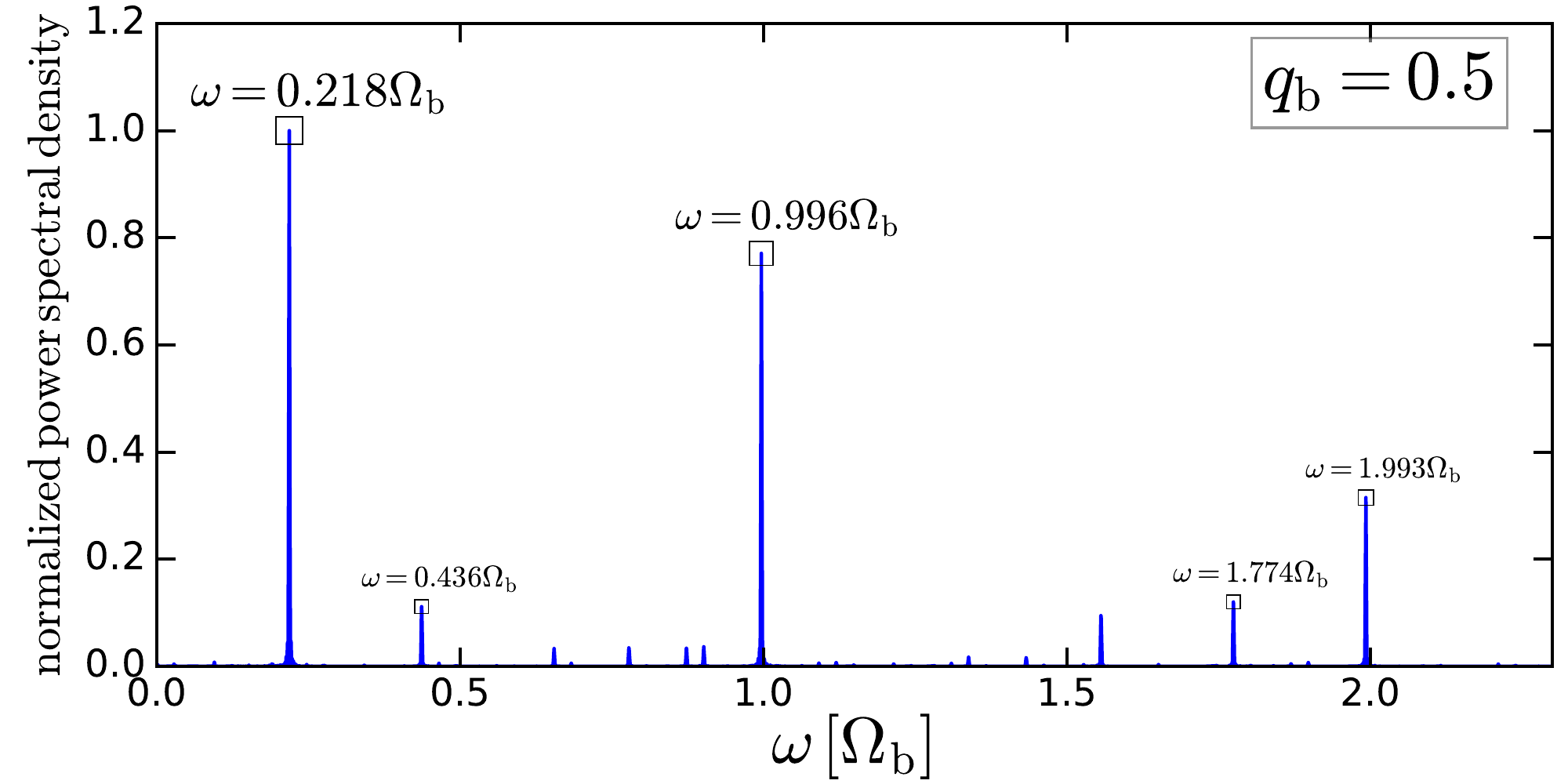}
\caption{
{Normalized power spectral density for the four simulations depicted in Figure~\ref{fig:accretion_rate_comparisons}.
For $q_{\rm b}\geq0.6$, the variability is clearly dominated by the frequency $\omega\sim \tfrac{1}{5}\Omega_{\rm b}$.
For $q_{\rm b}=0.5$, however, there is significant power at  $\omega\simeq\Omega_{\rm b}$. }
\label{fig:spectral_density}}
\end{figure*}

\begin{figure}
\includegraphics[width=0.45\textwidth]{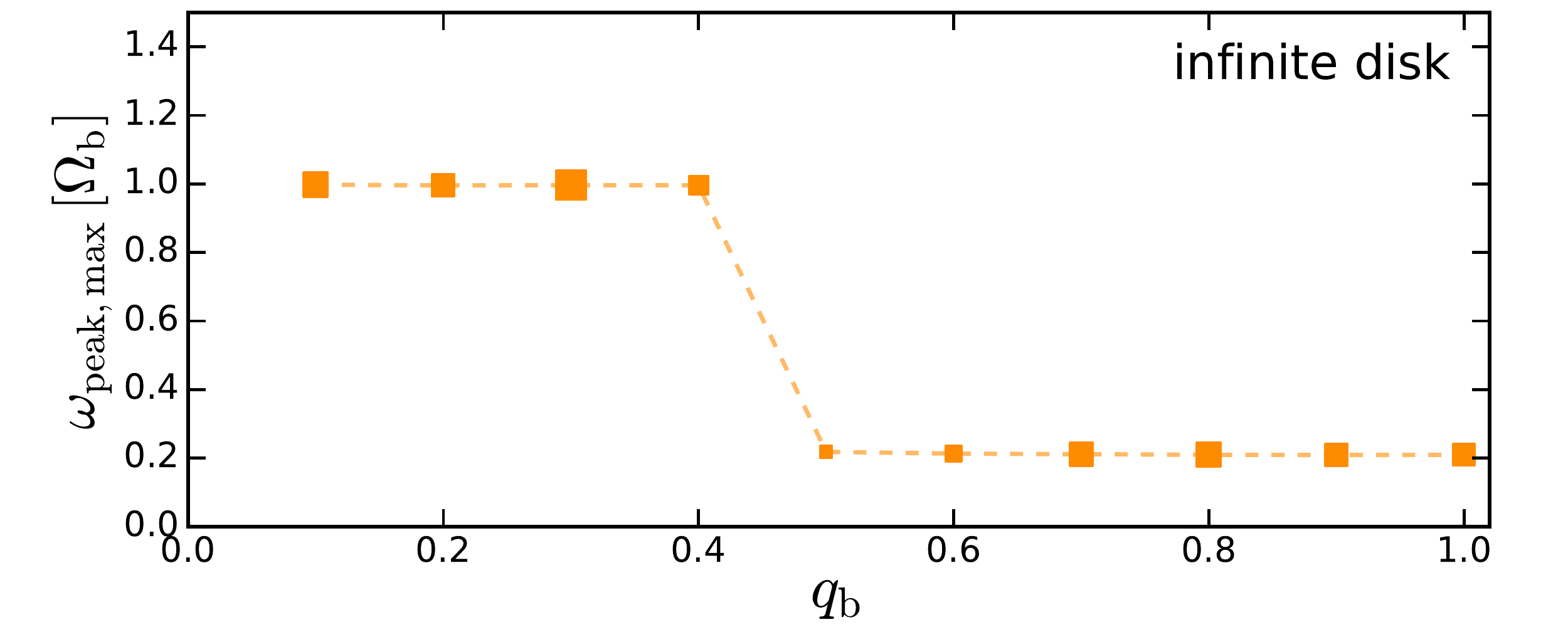}
\caption{
{
Dominant frequency $\omega_{\rm peak, max}$ obtained from the spectral analysis of $\dott{M}_{\rm b}$
for different values of $q_{\rm b}$ (Figure~\ref{fig:spectral_density}). The size
of the markers measures the relative power at this frequency relative to other peaks 
found in the PSD. The dominant frequency is $\sim \tfrac{1}{5}\Omega_{\rm b}$  for $q_{\rm b}\geq0.5$
although its power decreases with decreasing $q_{\rm b}$. For
$q_{\rm b}=0.4$, the accretion rate time series is dominated by the harmonic with 
$\omega_{\rm peak}\simeq\Omega_{\rm b}$
}
\label{fig:collected_frequencies}}
\end{figure}

The viscous time at a distance $R$ is
\begin{equation}
t_{\nu}
=\frac{4}{9}\frac{R^2}{\nu}
=\frac{2^{5/2}P_{\rm b}}{9\pi\alpha h_0^2}\left(\frac{R}{2a_{\rm b}}\right)^{3/2}~~,
\end{equation}
where $P_{\rm b}=2\pi/\Omega_{\rm b}$ is the orbital period of the binary.
With the fiducial values of $h_0=\alpha=0.1$, 
we find $t_\nu=200P_{\rm b}$ at $R=2a_{\rm b}$ (the edge of the cavity), 
which means that after a few hundred
binary periods, the system should begin to approach steady-state.  Figure~\ref{fig:image_comparisons} (top panel) shows the gas distribution in the vicinity
of a binary with $q_{\rm b}=0.5$ at four different times: $t=10,~100,~200,~750P_{\rm b}$. 
At $t=10 P_{\rm b}$, the cavity is still being filled up by viscous accretion from the CBD, and the CSDs are
beginning to form. After $t=200 P_{\rm b}$, the amount of mass in the CSDs is approximately constant, as is the mean surface density at the cavity edge. 

\begin{figure}
\includegraphics[width=0.45\textwidth]{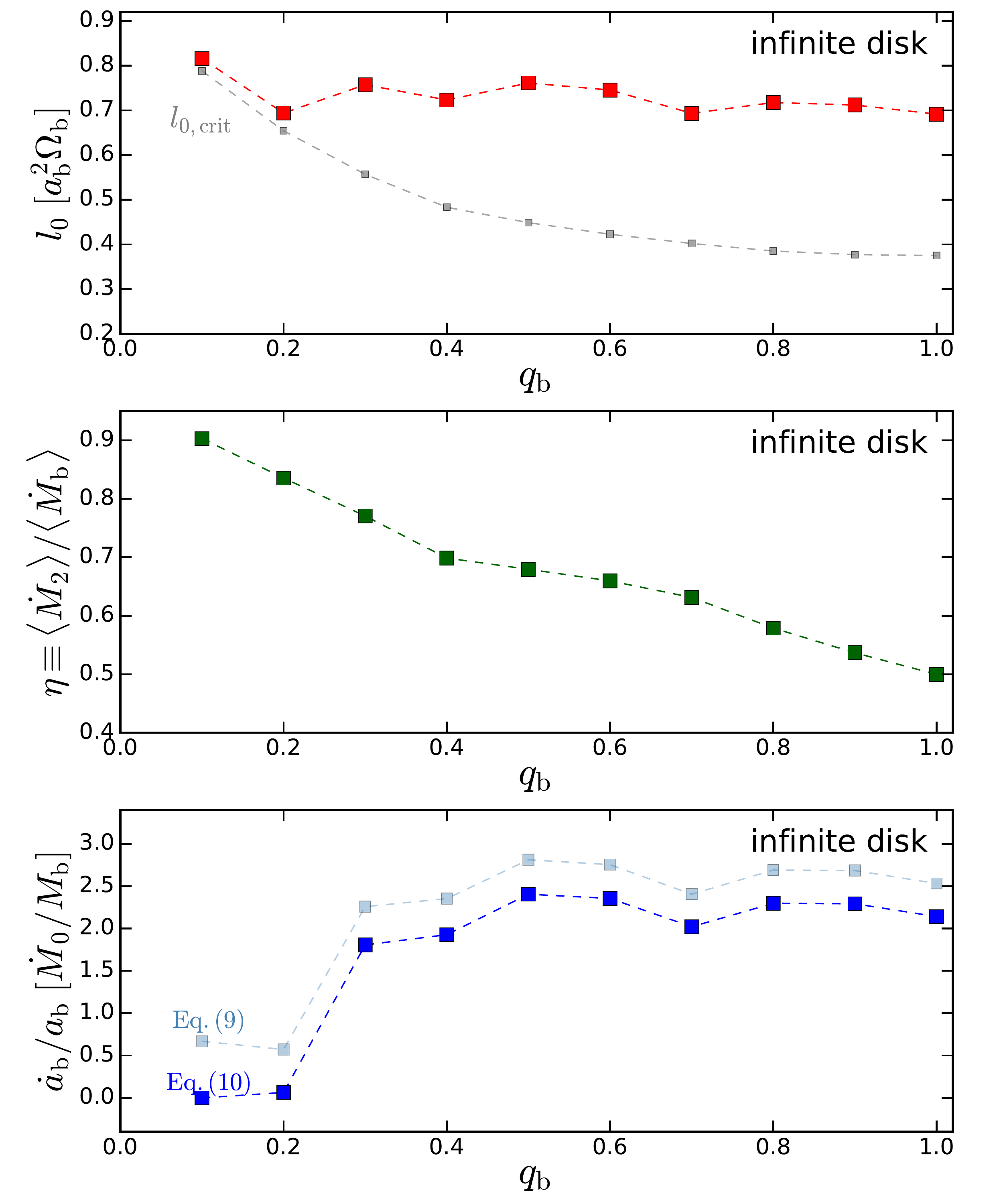}
\vspace{-0.1in}
\caption{Accretion eigenvalue $l_0$ (top), accretion rate ratio $\eta$ (middle) and binary migration rate
$\langle\dot{a}_{\rm b}\rangle /a_{\rm b}$
(bottom)  for a range of values in mass ratio $q_{\rm b}$ obtained from infinite disks in steady-state
 (see Fig.~\ref{fig:accretion_rate_comparisons}). As evidenced by the positive values
 of $\langle\dot{a}_{\rm b}\rangle$, binaries {with $q_{\rm b}\gtrsim0.3$ }expand while accreting.
\label{fig:collected_results}}
\end{figure}

\subsection{Quasi-Steady Accretion, Variability and Angular Momentum Transfer}
After a few thousands of binary orbits, the time-averaged accretion rate
 $\langle\dott{M}_{\rm b}\rangle$  matches the supply
rate $\dott{M}_0$ \citep{mun16b,mir17,mun19}. The top panel of
Fig.~\ref{fig:steady_accretion} shows $\dott{M}_{\rm b}$ for a $q_{\rm b}=0.4$ binary
as a function of time over 500 binary orbits. The bottom panel Fig.~\ref{fig:steady_accretion} 
shows $\langle\dott{M}_{\rm b}\rangle_{30}$,  the total accretion rate after performing a running average,
where the subscript denotes the width of the averaging window ($30P_{\rm b}$). Similarly,
$\langle\dott{M}_{1}\rangle_{30}$ and $\langle\dott{M}_{2}\rangle_{30}$ correspond to
the running averages for the primary and secondary accretion rates,
 respectively.
Once the high-frequency
variability is removed, these time series are constant, demonstrating the steady-state nature of our simulations.

We are interested in the steady-sate behavior of the angular momentum transfer rate
onto the binary 
$\dot{J}_{\rm b}$.  This is given by
\begin{equation}\label{eq:total_torque}
\dott{J}_{\rm b}=\dot{L}_{\rm b} + \dot{S}_1 +\dot{S}_2~~,
\end{equation}
where  $\dot{S}_{1,2}$ is the spin torque 
onto the primary/secondary and $\dot{L}_{\rm b}$ is the orbital angular momentum change rate.
Since $L_{\rm b}= \mu_{\rm b} l_{\rm b}$, where the reduced mass
 $\mu_{\rm b}=M_1M_2/M_{\rm b}=q_{\rm b}M_{\rm b}/(1+q_{\rm b})^2$
  and 
  $l_{\rm  b} = a_b^2\Omega_{\rm b}$ is the speciifc angular momemtum
  of a circular binary, we compute $\dott  L_{\rm b}$ via
\begin{equation}
\dott{L}_{\rm b}=\frac{1}{(1+q_{\rm b})^2}
\bigg[q_{\rm b}M_{\rm b}\frac{d l_{\rm b}}{dt}\bigg|_{\rm ext}
\!\!\!+\left(\dott{M}_2+q_{\rm b}^2\dott{M}_1\right)\,l_{\rm b}
\bigg]
\end{equation}
%
where the specific torque due to external forces  $dl_{\rm b}/dt|_{\rm ext}$
(including both gravity and accretion) 
is computed directly from simulation output, as are $\dott{M}_1$ and $\dott{M}_2$
 \citep[see][for more details]{mun19}.

Just like the binary accretion rate $\dott{M}_{\rm b}$, 
the transfer rate of angular momentum $\dot{J}_{\rm b}$
also reaches a stationary behavior without long-term trends. 
We show the stationary time series  in Fig.~\ref{fig:accretion_rate_comparisons}, 
where the normalized accretion rates onto the primary and secondary, 
$\dott{M}_1/\langle\dott{M}_{\rm b}\rangle$ and
 $\dott{M}_2/\langle\dott{M}_{\rm b}\rangle$ , and the 
normalized angular momentum transfer 
rate $\dot{J}_{\rm b}/\langle\dott{M}_{\rm b}\rangle$ are 
plotted as a function of time for different values of $q_{\rm b}$. 
The time-averaged accretion rates, 
$\langle \dott{M}_1\rangle$ and
$\langle \dott{M}_2\rangle$, differ from each other, with the lower-mass body receiving more mass
\citep[e.g.,][]{bat00,far14}. 

\subsubsection{Short-term Variability}

{
The stationarity of the time series in Figure~\ref{fig:accretion_rate_comparisons} allows us the carry out
spectral analysis of $\dott{M}_{\rm b}$. Using the Lomb-Scarle periodogram, we compute
the power spectral density (PSD, in units of $\dott{M}_0^2\Omega_{\rm b}$) for each of the $\dot{M}_{\rm b}$ 
time series in Figure~\ref{fig:accretion_rate_comparisons}, 
and normalize them by their peak values. These normalized PSDs are shown in Figure~\ref{fig:spectral_density}.
For large mass ratios, the dominant frequency is $\omega_{\rm peak,max}\sim \tfrac{1}{5}\Omega_{\rm b}$
as has been noticed by several previous studies \citep{mac08,shi12,far14,mun16b,mir17},
although the earlier works prior to 2016
did not reach quasi-steady state in the simulations.
The other peaks in the spectral
density are attributed to harmonics of $\tfrac{1}{5}\Omega_{\rm b}$ and to the binary orbital
frequency $\Omega_{\rm b}$ and its harmonics. For $q_{\rm b}\lesssim0.6$ (not shown in the figure),
the frequency $\simeq\Omega_{\rm b}$ sits at $20\%$ of the maximum power; for $q_{\rm b}=0.5$
it sits at nearly $80\%$ of the maximum power; when $q_{\rm b}=0.4$, the maximum power shifts
to $\omega_{\rm peak,max}\simeq\Omega_{\rm b}$ (see Figure~\ref{fig:collected_frequencies} ).
This switch in the fundamental frequency from $\tfrac{1}{5}\Omega_{\rm b}$ to  $\Omega_{\rm b}$ 
at $q_{\rm b}\lesssim 0.4$ is roughly consistent with the spectral analysis of \citet{far14}.
}

\subsubsection{Secular Behavior}
The accretion "eigenvalue", defined by 
\begin{equation}
l_0\equiv\frac{\langle\dot{J}_{\rm b}\rangle}{\langle\dott{M}_{\rm b}\rangle}~~,
\end{equation}
gives the angular momentum received by the binary per unit of accreted mass.
We show the numerical results of $l_0$ for a range of values of $q_{\rm b}$
in the top panel of Fig.~\ref{fig:collected_results};
the values of $l_0$ lie in the range $[0.65,0.85]a_{\rm b}^2\Omega_{\rm b}$ for all mass ratios explored. 
As in \citet{mun19}, we find that the time-averaged spin torques
$\langle \dot{S}_1\rangle$ and
$\langle \dot{S}_2\rangle$ are much smaller than $\langle \dot{L}_{\rm b}\rangle$
(the contribution of $\langle\dot{S}_{1,2}\rangle$
to $\langle \dott{J}_{\rm b}\rangle$
 is $3\%-5\%$ for $r_{\rm acc}=0.03a_{\rm b}$).

For circular binaries, the orbital angular momentum change can be written as
\begin{equation}
\frac{\dot{L}_{\rm b}}{L_{\rm b}}
=\frac{\dott{M}_1}{M_1}+\frac{\dott{M}_2}{M_2}
-\frac{1}{2}\frac{\dott{M}_{\rm b}}{M_{\rm b}}
+\frac{1}{2}\frac{\dot{a}_{\rm b}}{a_{\rm b}}
\end{equation}
Since $\langle \dot{L}_{\rm b}\rangle\simeq\langle \dot{J}_{\rm b}\rangle$,  we find that 
the eigenvalue $l_0$ and the secular migration rate $\langle \dot{a}_{\rm b}\rangle$ are
 related via
\begin{equation}\label{eq:migration_rate_ideal}
\begin{split}
\frac{\langle\dot{a}_{\rm b}\rangle}{a_{\rm b}}=&
2\frac{\langle\dott{M}_{\rm b}\rangle}{M_{\rm b}}
\!\bigg\{\frac{(1+q_{\rm b})^2}{q_{\rm b}}\frac{l_0}{a_{\rm b}^2\Omega_{\rm b}}
-(1-\eta)(1+q_{\rm b})
-\eta \frac{1+q_{\rm b}}{q_{\rm b}}
+\frac{1}{2}
\!\!\bigg\}\\=&
2\frac{\langle\dott{M}_{\rm b}\rangle}{M_{\rm b}}
\frac{(1+q_{\rm b})^2}{q_{\rm b}}\frac{l_0-l_{0,{\rm crit}}}{a_{\rm b}^2\Omega_{\rm b}}
\end{split}
\end{equation}
where we have defined the ``preferential accretion rate ratio''
$\eta=\langle\dott{M}_2\rangle/\langle\dott{M}_{\rm b}\rangle$.
Note from Equation~(\ref{eq:migration_rate_ideal}) that $l_0$
needs to be greater than a critical value  $l_{0,{\rm crit}}$ to result in $\langle \dot{a}_{\rm b}\rangle>0$~.
This threshold quantity is also  shown in the top panel of Fig.~\ref{fig:collected_results}.
We see that that $l_0$ is significantly above the threshold that leads to binary expansion
{for $q_{\rm b}\gtrsim0.3$. When $q_{\rm b}\lesssim0.2$, the value of $l_0$ is very close
to $l_{0,{\rm crit}}$, because $\eta$ is larger.}
{In Fig.~\ref{fig:collected_results}, we also show the values of $\eta$ (middle panel) as a function
of $q_{\rm b}$. If $l_0\lesssim l_{0,{\rm crit}}$, then inward migration is possible, even if
$l_0$ is positive.}

We can measure the migration rate directly from  the instantaneous change in the
binary's specific orbital energy (${\cal E}_{\rm b}=-\tfrac{1}{2}{\cal G}M_{\rm b}/a_{\rm b}$):
\begin{equation}\label{eq:migration_rate}
\frac{\dot{a}_{\rm b}}{a_{\rm b}}=-\frac{\dott{{\cal E}}_{\rm b}}{{\cal E}_{\rm b}}+\frac{\dott{M}_{\rm b}}{M_{\rm b}}~~.
\end{equation}
The change in orbital energy is 
$\dott{{\cal E}}_{\rm b} =-{\mathcal{G}\dott{M}_{\rm b}}/|{\bf r}_{\rm b}|+
\dot{\mathbf{r}}_{\rm b}\cdot \mathbf{f}_{\rm ext}$~, where  ${\bf r}_{\rm b}$ 
is the binary separation vector 
\citep[][eq.~33]{mun19}, and the external forces $ \mathbf{f}_{\rm ext}$ include
both gravity and accretion kicks, which
are computed on-the-fly for each time-step of the simulation.

The secular migration rate $\langle \dot{a}_{\rm b}\rangle /a_{\rm b}$ (Equation~\ref{eq:migration_rate})
is shown as a function
of $q_{\rm b}$ in the bottom panel of Fig.~\ref{fig:collected_results}. For comparison, we also
include the migration rate computed from Equation~\ref{eq:migration_rate_ideal}.
As implied by the 
corresponding values of $l_0$, binaries with  $q_{\rm b}\gtrsim0.2$ exhibit positive migration rates, \
{those $q_{\rm b}\lesssim0.2$ exhibit a sharp drop in the migration rate down to $\dot{a}_{\rm b}\approx0$.}
These results extend the findings of \citet{mun19} to binaries of different mass ratios.

\section{Accretion from Finite  ``Tori''}\label{sec:finite}
The simulations presented in Section~\ref{sec:infinite}
pertain to "infinite" disks with a constant supply rate 
$\dott  M_0$. 
 Realistic systems may not have a constant gas supply and a true steady state 
 -- in a strict sense -- is not possible.
Nevertheless, if the supply rate changes {\it slowly} (e.g., by a CBD that is running out of mass), 
this quantity may still act as a scaling parameter
that is being ``adiabatically''  dialed down/up. 
In such case, the amount of angular momentum transferred to the binary would decrease/grow in proportion to
the supply rate, but the angular momentum transferred {\it per unit accreted mass} -- i.e., $l_0$ -- would remain unchanged.
To test whether this is indeed the case, here we consider finite disks/tori, 
in which the supply rate onto the binary is self-consistently set by the viscous evolution of a 
finite reservoir of mass.

\begin{figure}
\includegraphics[width=0.45\textwidth]{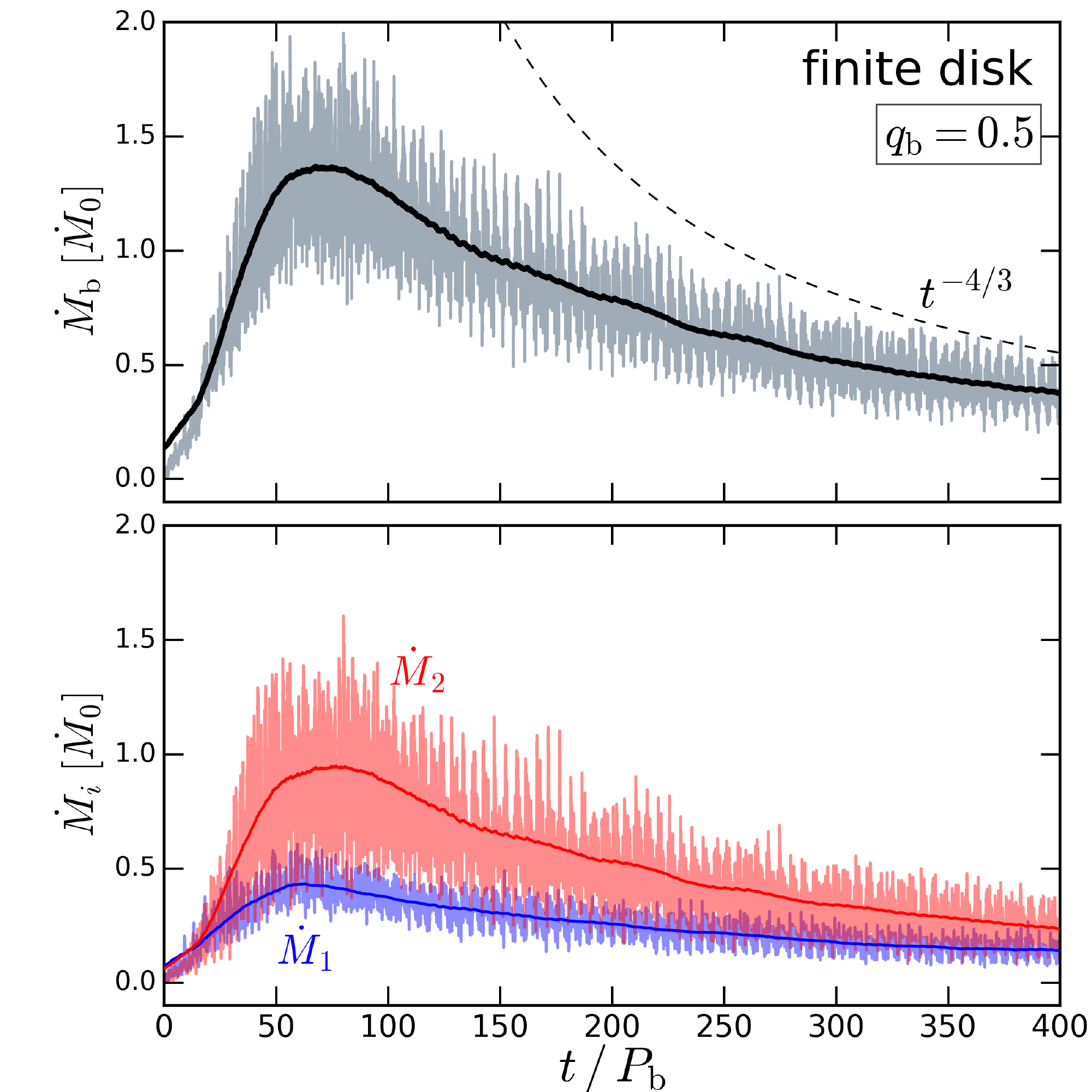}
\includegraphics[width=0.43\textwidth]{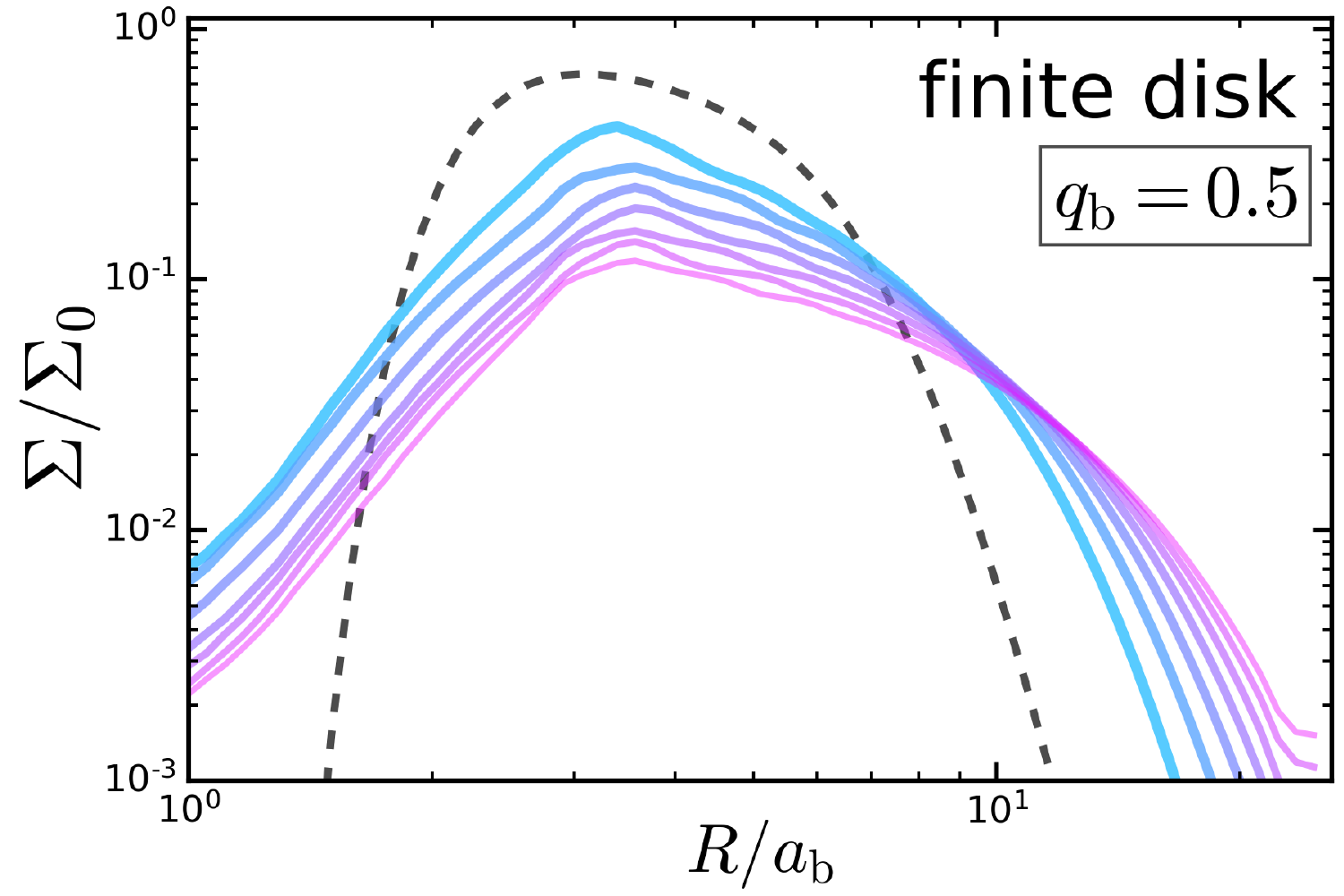}
\caption{Accretion and depletion of a finite circumbinary torus. The top two panels show evolution of the total binary accretion
rate $\dott{M}_{\rm b}$ (black lines; top) 
and the individual accretion rates $\dott{M}_1$  and $\dott{M}_2$
as a function of time.
Accretion rates are scaled by $\dott{M}_0=3\pi \alpha h_0^2\Sigma_0 a_{\rm b}^2\Omega_{\rm b}$.
Solid thick lines depict the running averages, $\langle \dott{M}_{\rm b}\rangle_{30}$ $\langle \dott{M}_1\rangle_{30}$
and $\langle \dott{M}_1\rangle_{30}$ , where $\langle\cdot\rangle_{30}$ denotes time-averaging with a running window of
30 binary orbits (see text).
The bottom panel shows the evolution of the azimuthally-averaged surface density profile $\Sigma(R)$ in time.
Curves are spaced in intervals of $60 P_{\rm b}$ (from $t=63P_{\rm b}$ in cyan to  $t=483P_{\rm b}$ in magenta)
and each consists of the time-averaged profile over 6 binary orbits. The dashed black line depics 
the initial condition (Equation~\ref{eq:initial_profile_finite}).
\label{fig:torus_evolution}}
\end{figure}

To model an accretion torus, we modify the initial density profile
(Equation~\ref{eq:initial_profile_infinite}), by 
multiplying by an exponential tapering function, i.e.,
\begin{equation}\label{eq:initial_profile_finite}
{\Sigma}(R)\rightarrow{\Sigma}(R)
\!\!\left[1+\exp{(R-R_{\rm disk})}\right]^{-1}~~,
\end{equation}
where the  term in square brackets forces the gas
density to drop exponentially away from the central binary. The density profile
(\ref{eq:initial_profile_finite})
 peaks at $R\lesssim R_{\rm disk}$, and we set  $R_{\rm disk}=6a_{\rm b}$. The red curve in Fig~\ref{fig:initial_density_profiles}
depicts this modified density profile, highlighting the outer region of the disk that will spread due to viscous stresses \citep[e.g.,][]{lyn74,har98}.

The simulation setup is analogous to the one detailed in Section~\ref{sec:methods}
except for the outer boundary condition:  we choose the computational domain to be a square 
 of side $65a_{\rm b}$ with a background
density of $10^{-10}\Sigma_0$ into which the disk is allowed to viscously expand; that is, the outer boundary condition of a viscously expanding disk is trivially 
captured by the quasi-Lagrangian nature of the {\footnotesize AREPO} scheme.

\begin{figure}
\centering
\includegraphics[width=0.41\textwidth]{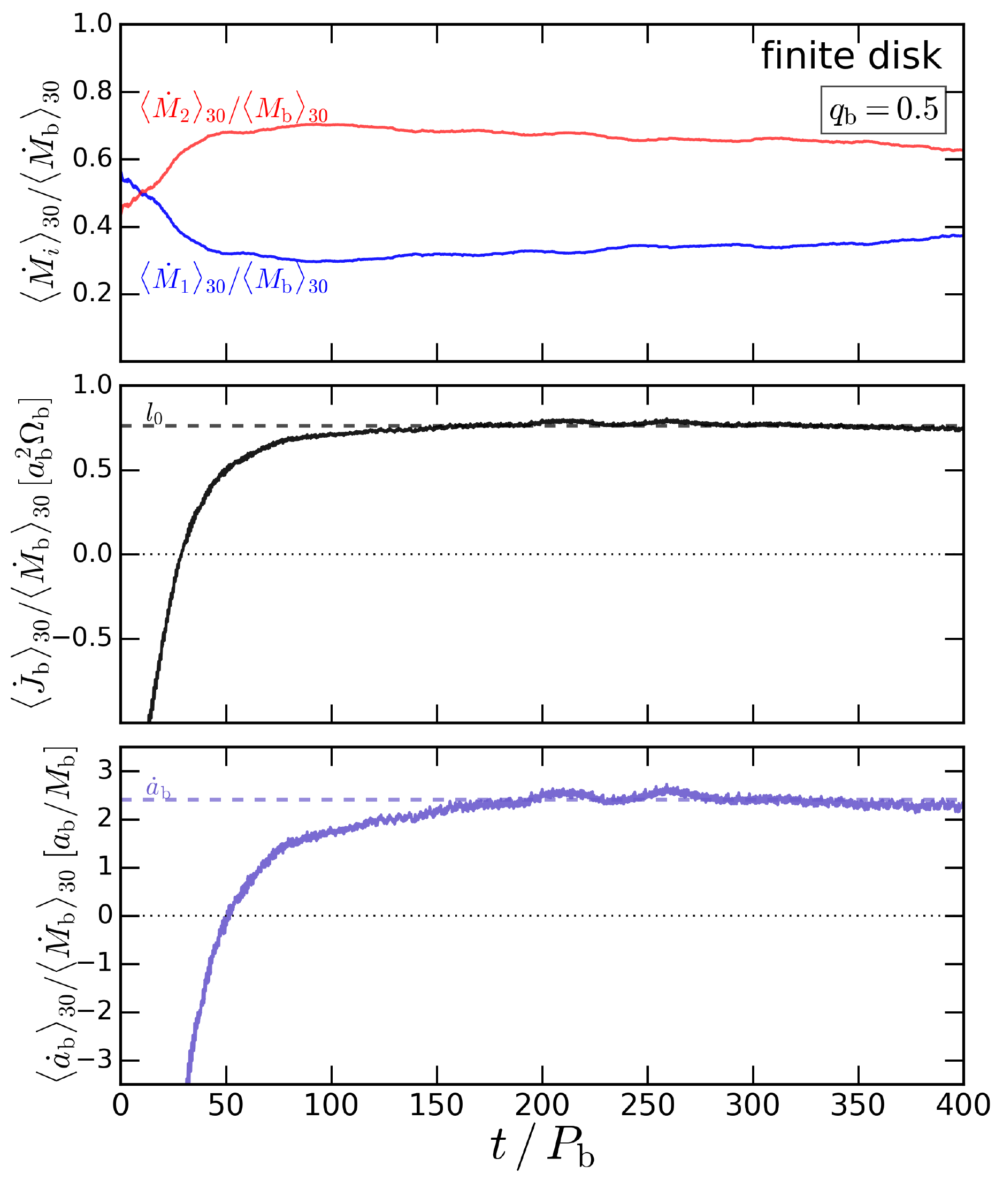}
\caption{
Transient and pseudo-stationary (i.e., with slowly-evolving secular trends) accretion from a finite torus
with $q_{\rm b}=0.5$.
The top panel depicts the total binary accretion rate $\langle\dott{M}_1\rangle_{30}$ (blue) and
$\langle\dott{M}_2\rangle_{30}$ (red) normalized by the running-average net accretion rate
$\langle\dott{M}_{\rm b}\rangle_{30}$. The middle panel depicts the running average of
the angular momentum transfer rate $\langle\dot{J}_{\rm b}\rangle_{30}$ normalized by
$\langle\dott{M}_{\rm b}\rangle_{30}$. The bottom panel depicts the running average of
the migration rate $\langle\dot{a}_{\rm b}\rangle_{30}$  normalized by
$\langle\dott{M}_{\rm b}\rangle_{30}$. The reference eigenvalue $l_0$ and migration
rate $\dot{a}_{\rm b}$ obtained for infinite disks (Fig.~\ref{fig:collected_results}) are represented by horizontal dashed lines.
During the transient phase ($t\lesssim 100P_{\rm b}$) -- which corresponds to the filling  of the initially 
empty central cavity -- both $\langle\dot{J}_{\rm b}\rangle_{30}$ and $\langle\dot{a}_{\rm b}\rangle_{30}$ are
 negative; this phase is followed by a pseudo-stationary state 
during which $\langle\dot{J}_{\rm b}\rangle_{30}$ and $\langle\dot{a}_{\rm b}\rangle_{30}$ 
 are similar to the infinite disk values when scaled by $\langle\dott{M}_{\rm b}\rangle_{30}$
\label{fig:finite_disk_accretion_1}}
\end{figure}

\subsection{Viscous Evolution of Finite Tori}\label{sec:viscous_spreading}
Due to viscous spreading, 
the supply rate onto the central binary eventually decreases.
We provide an example of such evolution in the lower panels of
Fig.~\ref{fig:image_comparisons}. The initial transient phase (the filling of the cavity, $t=10P_{\rm b}$) matches that of the ``infinite'' disk
case (shown in the upper panels). The CSDs are progressively filled until they reach a maximum mass ($t=100P_{\rm b}$). Once the transient phase ends, the 
circumbinary cavity can only be replenished by a CBD with ever-decreasing mass, resulting in a gas morphology that closely resembles
that of the ``infinite'' disk case ($t=200P_{\rm b}$ and $t=750P_{\rm b}$) but at a uniformly reduced density scale (compare upper and lower panels).
Eventually ($t\sim1000P_{\rm b}$),
the circumbinary torus becomes severely depleted, and the accretion rate onto the binary drops to a few percent of its peak value at earlier times.

\begin{figure}
\centering
\includegraphics[width=0.41\textwidth]{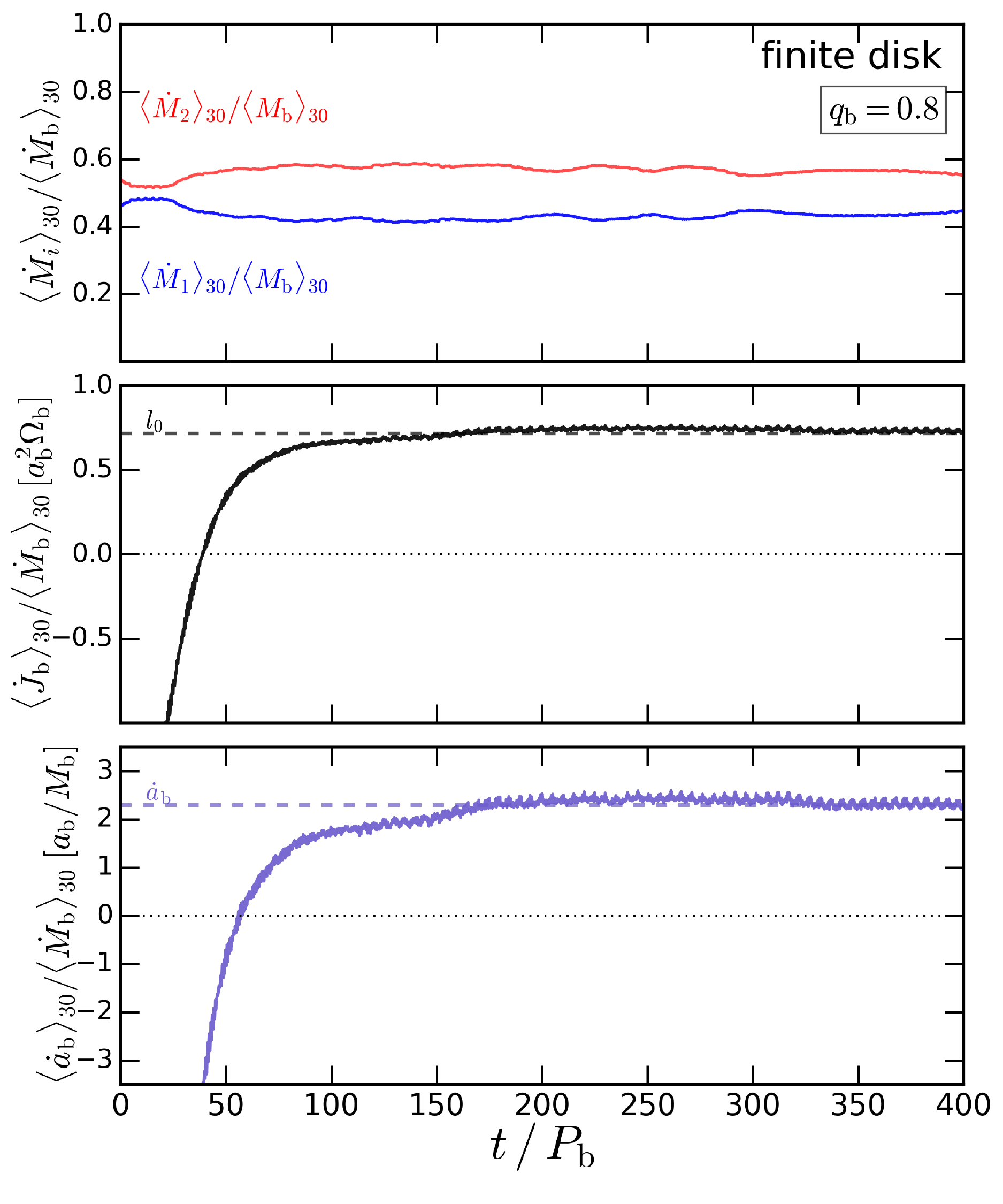}
\caption{
Same as Fig.~\ref{fig:finite_disk_accretion_1} but for $q_{\rm b}=0.8$.
\label{fig:finite_disk_accretion_2}}
\end{figure}

\begin{figure}
\centering
\includegraphics[width=0.41\textwidth]{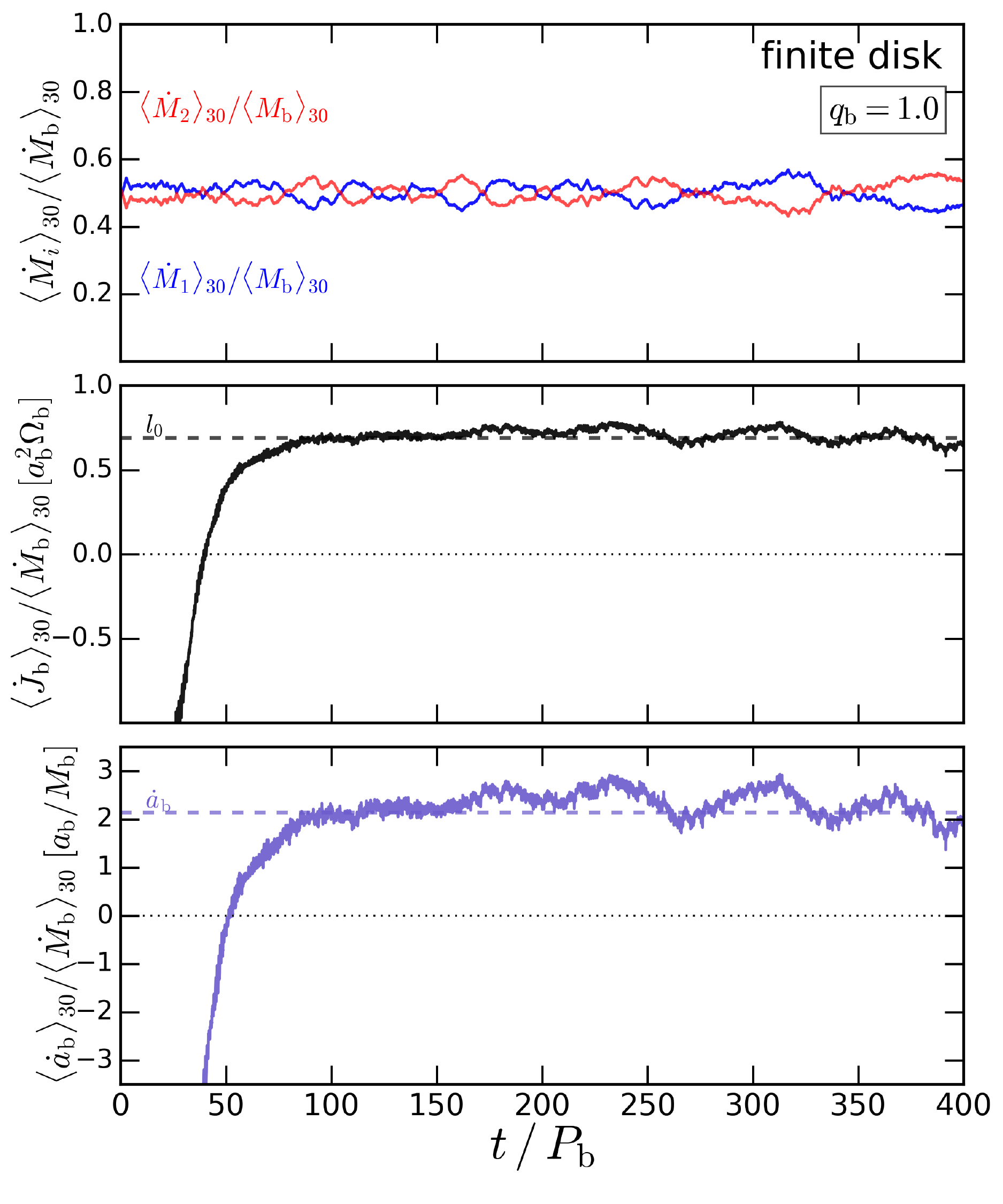}
\vspace{-0.05in}
\caption{
Same as Fig.~\ref{fig:finite_disk_accretion_1} but for $q_{\rm b}=1$.
\label{fig:finite_disk_accretion_3}}
\vspace{-0.1in}
\end{figure}

\begin{figure*}
\centering
\includegraphics[width=0.41\textwidth]{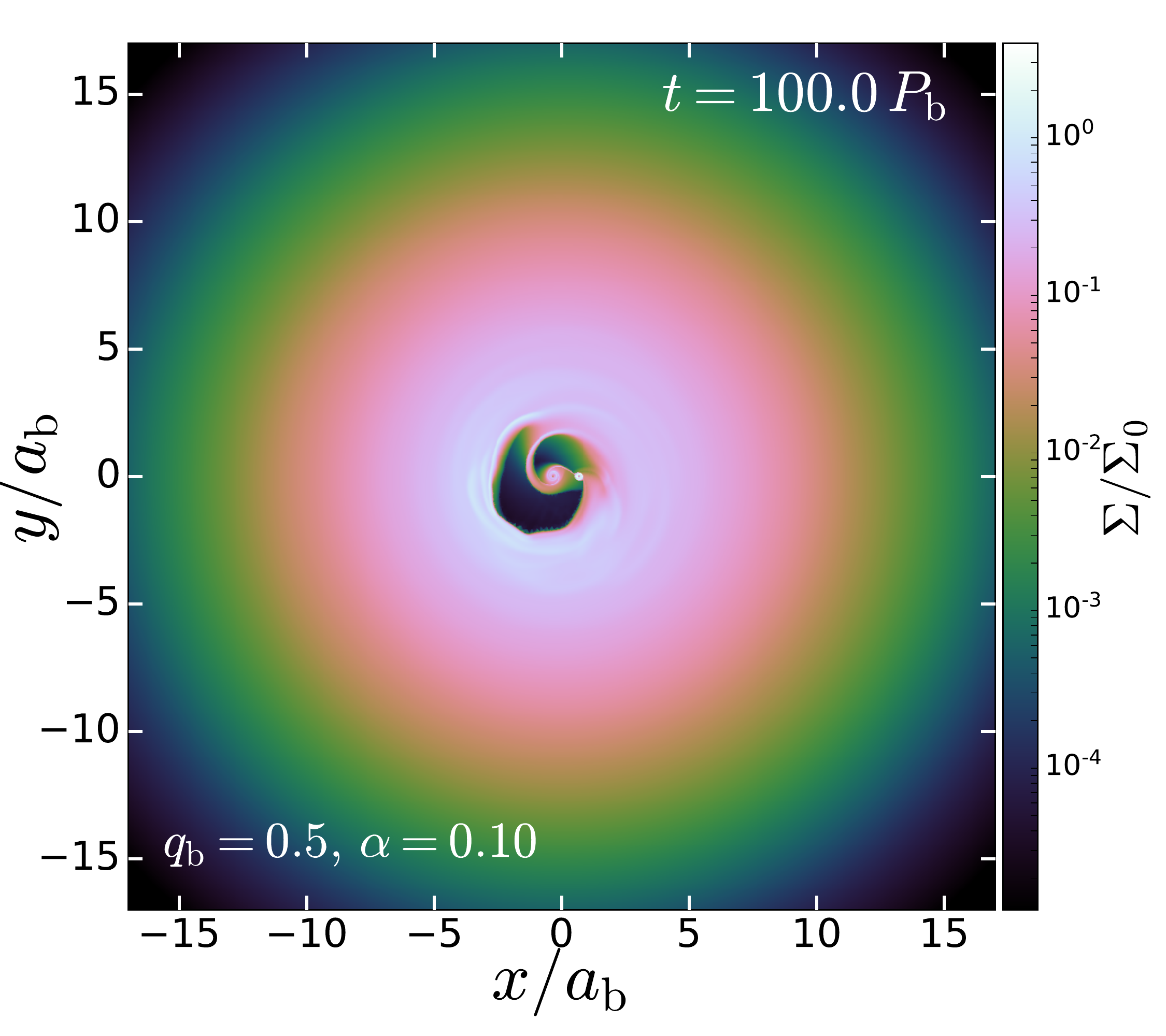}
\includegraphics[width=0.41\textwidth]{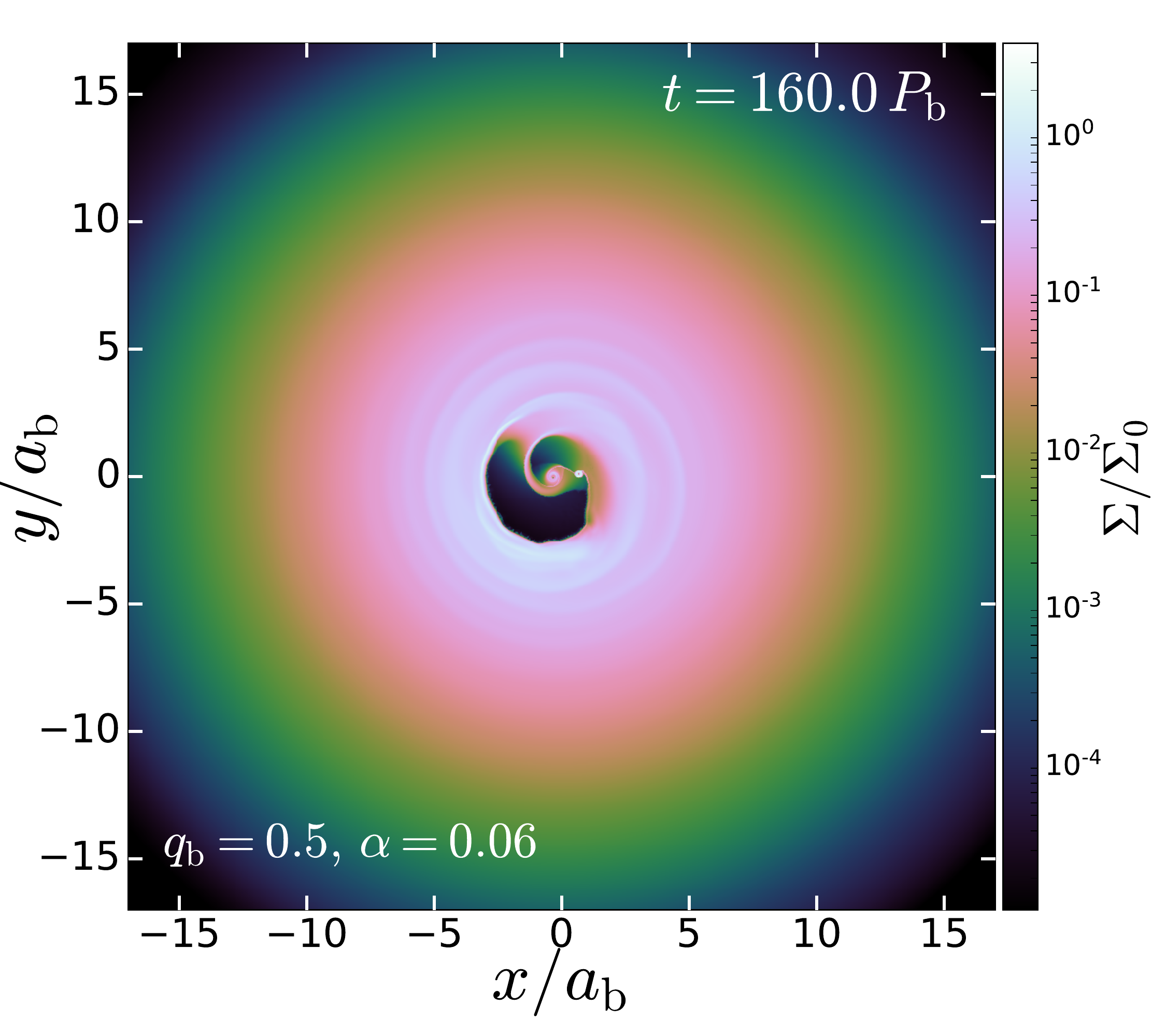}
\includegraphics[width=0.41\textwidth]{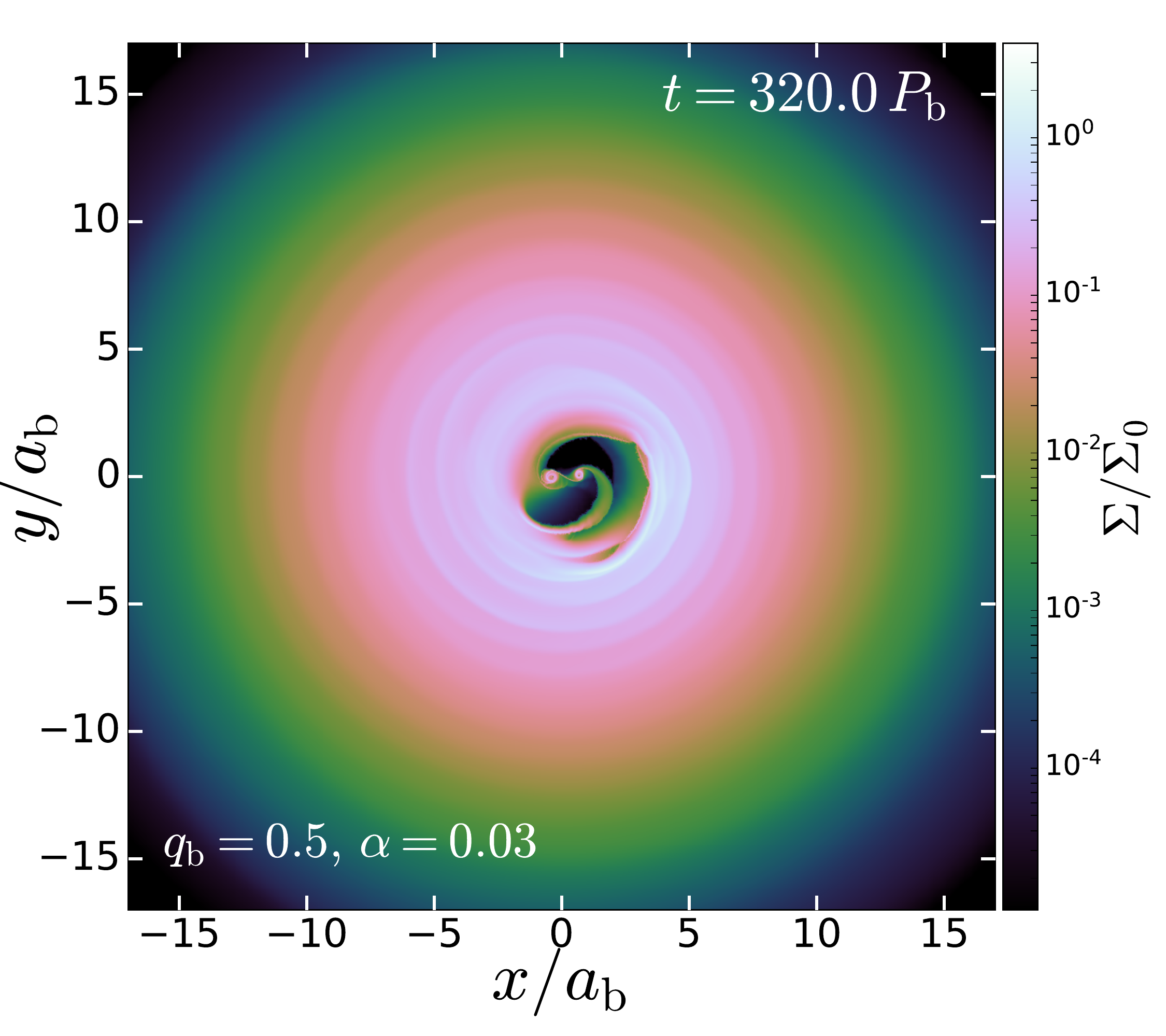}
\includegraphics[width=0.41\textwidth]{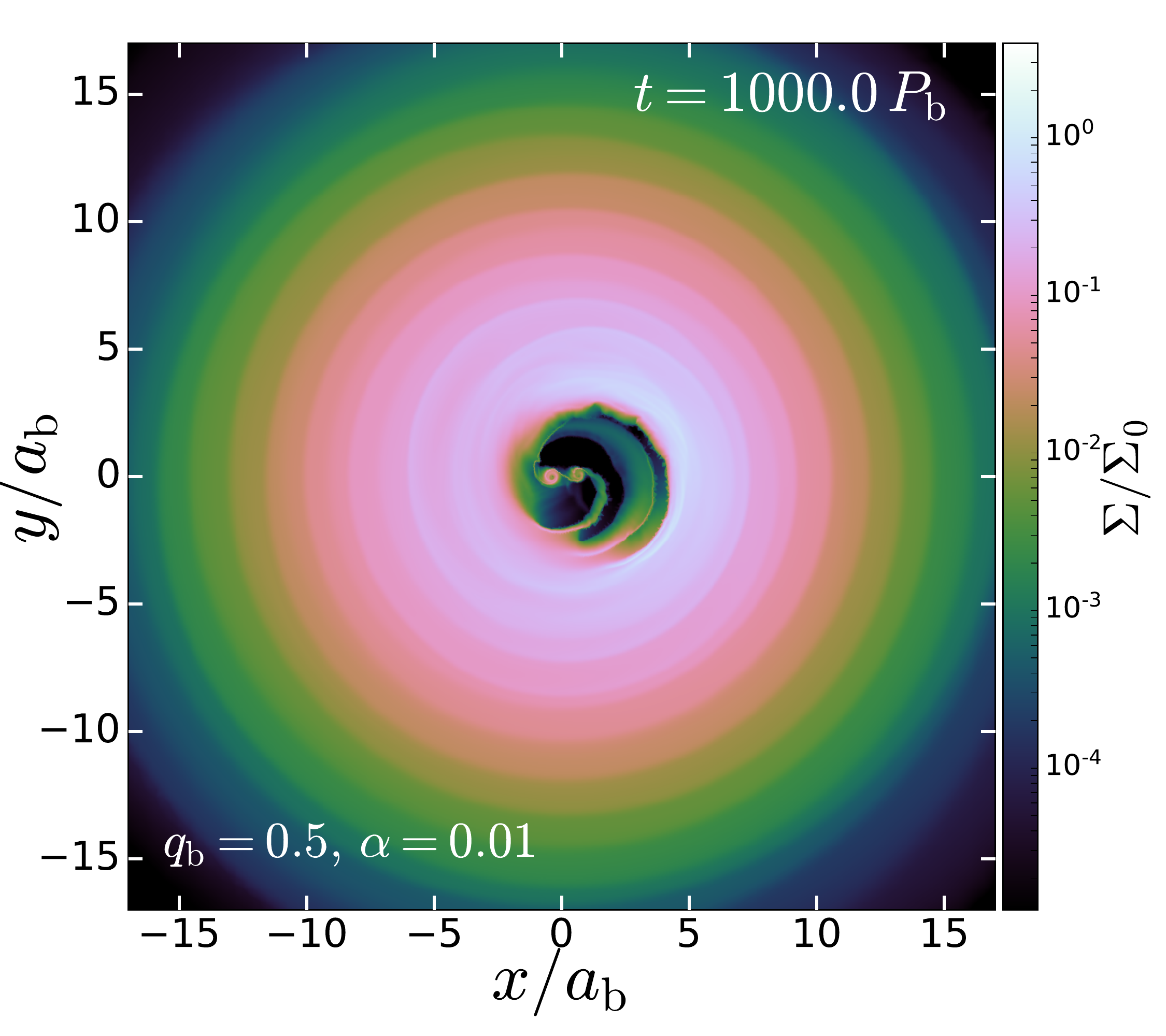}
\caption{Surface density field for accreting
tori of different viscosities around a binary with $q_{\rm b}=0.5$. From left to right, $\alpha=0.1$, 0.06, 0.03 and 0.01.
The different times in units of $P_{\rm b}$ correspond to roughly the same 
 dimensionless time ${\cal T}\approx 1.5$ (Equation~\ref{eq:dimensionless_time}) in all panels. The density scale
 $\Sigma_0$ (see. Equation~\ref{eq:density_scaling}) is different for different values of $\alpha$.
 {As in Figure~\ref{fig:image_comparisons},  the secondary is to the right of the primary in all panels.}
\label{fig:viscosity_comparison}}
\end{figure*}

\begin{figure}
\centering
\includegraphics[width=0.45\textwidth]{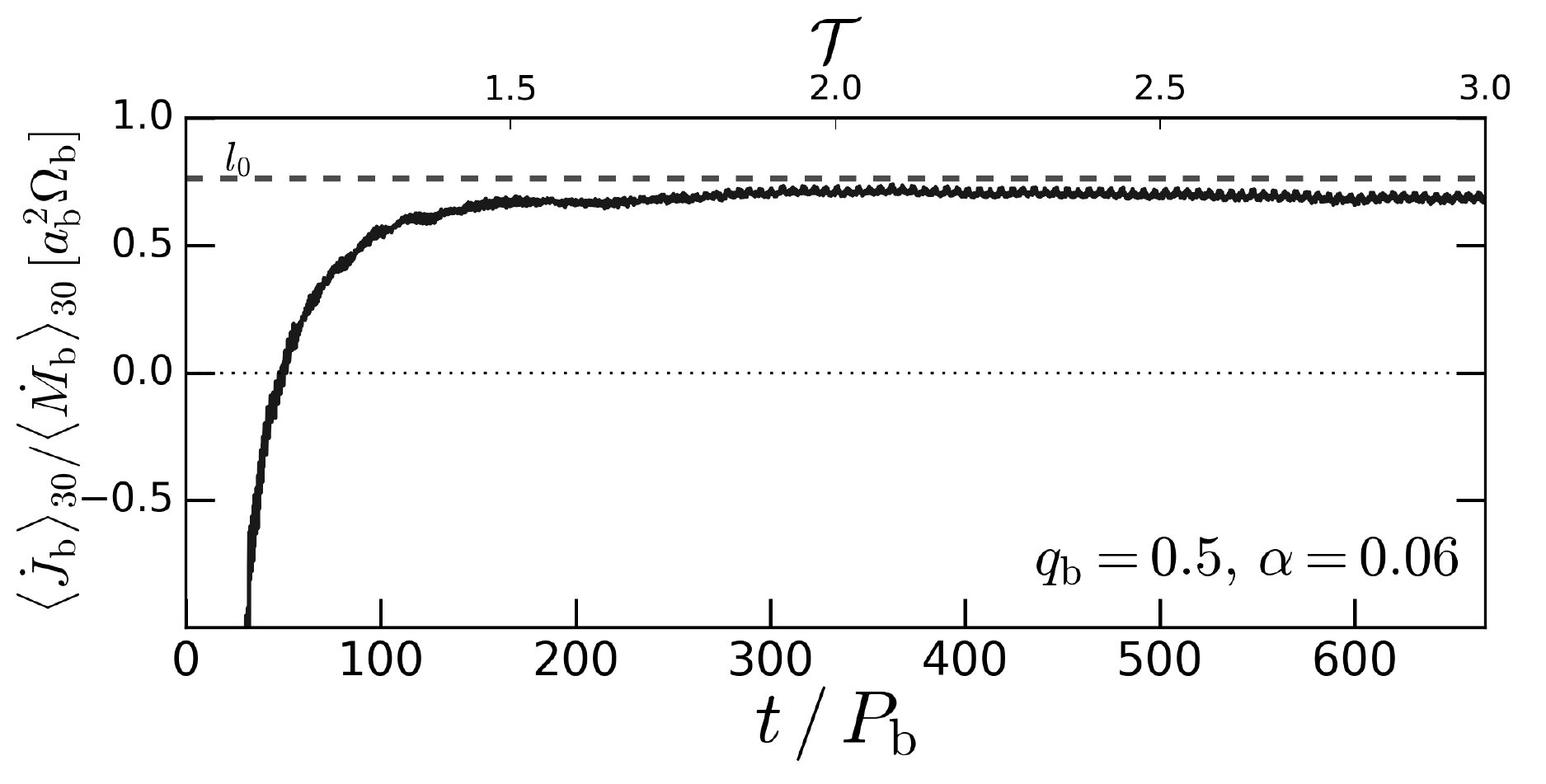}
\includegraphics[width=0.45\textwidth]{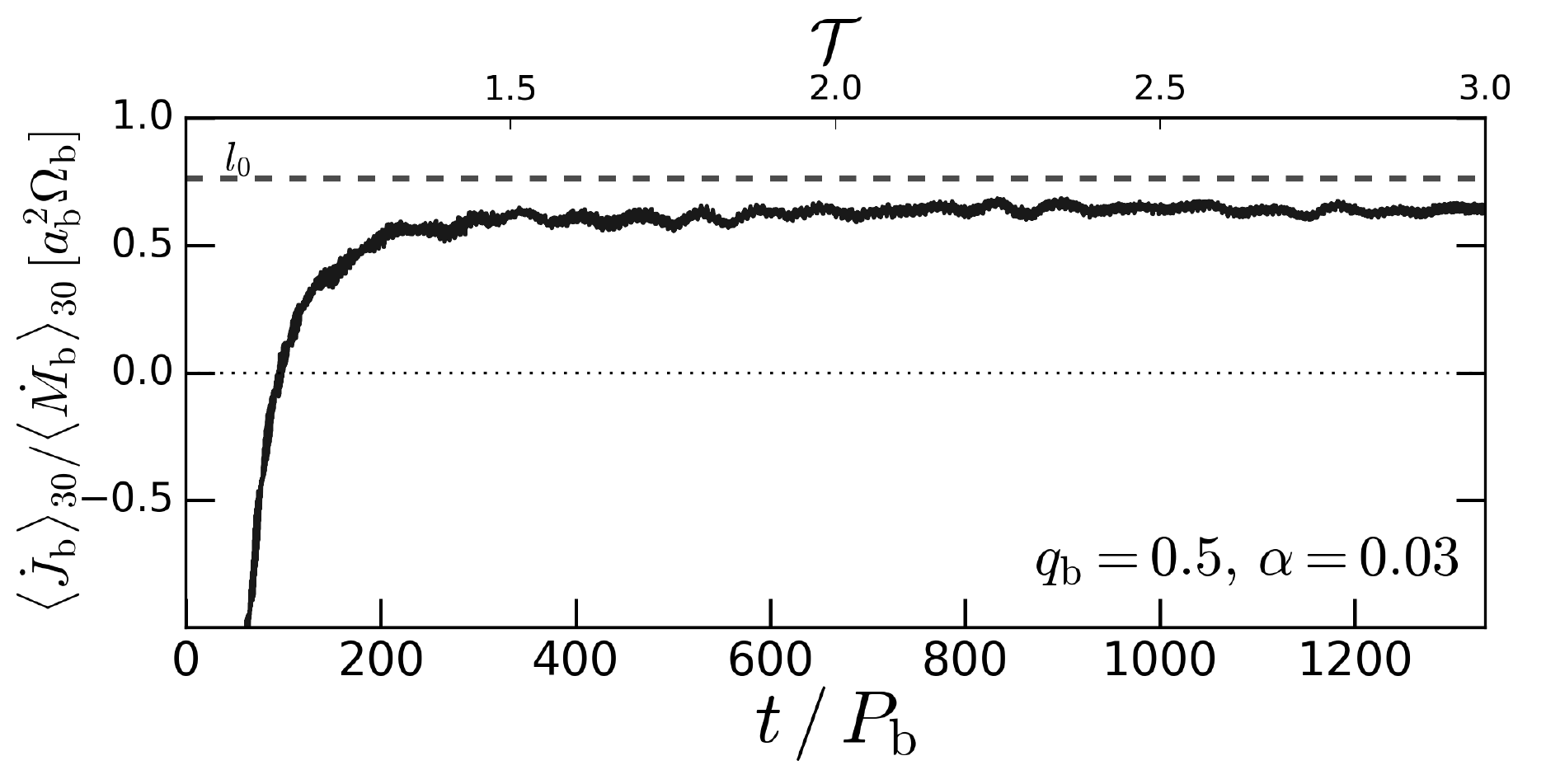}
\includegraphics[width=0.45\textwidth]{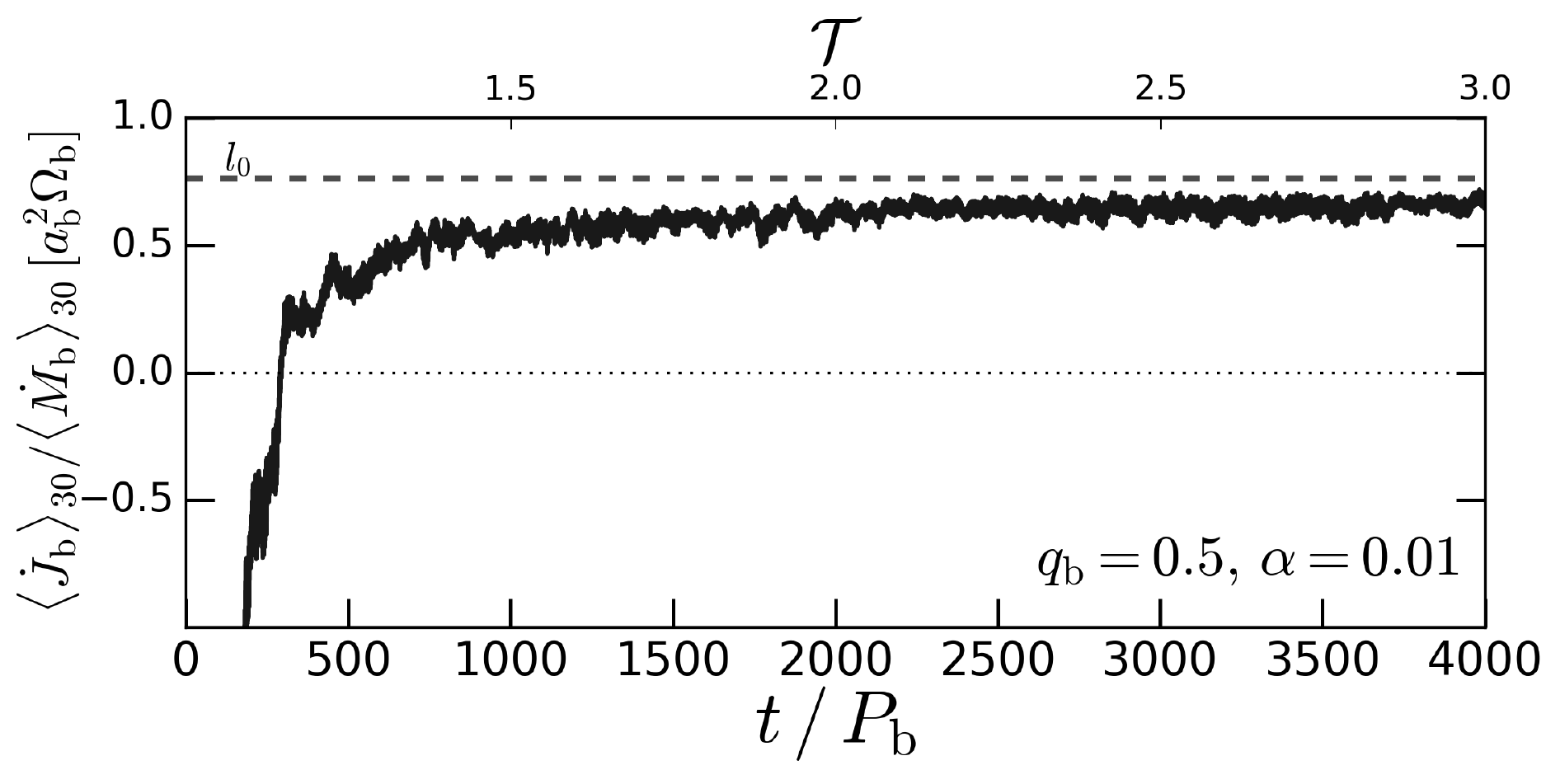}
\vspace{-0.1in}
\caption{Evolution of the normalized angular momentum transfer rate
 $\langle\dot{J}_{\rm b}\rangle_{30}/\langle\dott{M}_{\rm b}\rangle_{30}$ as a function of time for $q_{\rm b}=0.5$ and different
 values of the viscosity coefficient $\alpha$. From top to bottom:  $\alpha=0.06$, 0.03 and 0.01. In each panel, the bottom
 $x$-axis represents time in units of the binary orbital period $P_{\rm b}$, while the top $x$-axis represents time in dimensionless
 units  (see Equation~\ref{eq:density_scaling}). The dashed line indicates the reference eigenvalue $l_0$ obtained
 for ``infinite'' disks {with $\alpha=0.1$}.
\label{fig:viscosity_comparison_eigenvalue}}
\end{figure}

We show the evolution of an accreting torus in Figure~\ref{fig:torus_evolution}. During the transient phase ($t\lesssim100P_{\rm b}$), the accretion rate
rises rapidly as the cavity is filled; later on  ($t\gtrsim100P_{\rm b}$), the accretion
 rate decreases steadily (top panel).
A clearer picture of the long term trend is provided by the running average $\langle \dott{M}_{\rm b}\rangle_{30}$ (thicker line).
At later times,  $\langle \dott{M}_{\rm b}\rangle_{30}$ can be compared to the accretion 
rate onto a central point mass $\dott{M}_{\rm c}$  from a viscously evolving
disk of initial mass $M_{{\rm d},0}$ and initial characteristic radius $R_{{\rm d},0}$ \citep{har98,and09}
\begin{equation}
\dott{M}_{\rm c}=
\frac{3(2-\gamma)\nu_0}{2}\left(\frac{M_{{\rm d},0}}{R_{{\rm d},0}^2}\right)
\left(1+\frac{t}{t_{\nu,0}}\right)^{\tfrac{-5+2\gamma}{4-2\gamma}}
\end{equation}
where the prescribed viscosity profile is $\nu=\nu_0(R/R_{{\rm d},0})^\gamma$ and $t_{\nu,0}$ is the viscous
time at $R=R_{{\rm d},0}$. At later times, and for $\gamma{=}1/2$, 
we have $\dott{M}_{\rm c}\propto t^{-4/3}$, which is in rough agreement with the evolution of 
$\langle \dott{M}_{\rm b}\rangle_{30}$. 
In the middle panel of Fig.~\ref{fig:torus_evolution},
 we show the accretion rate onto the primary 
 $\dott{M}_1$ (in blue) and secondary $\dott{M}_2$ (in red) and their corresponding running averages 
$\langle \dott{M}_1\rangle_{30}$  and $\langle \dott{M}_2\rangle_{30}$, respectively (thicker lines).
As the net accretion rate 
$\dott{M}_{\rm b}$ diminishes, so do $\dott{M}_1$ and $\dott{M}_2$, although the dominance
of accretion onto the secondary is preserved throughout the duration of the simulation. The bottom panel of 
Fig.~\ref{fig:torus_evolution} depicts the density profile $\Sigma(R)$ of the torus at different times. When $t\sim500P_{\rm b}$,
the peak surface density has decreased by a factor of $\sim7$ and the radial extent has spread by a factor of $\sim2.5$.

\subsection{Angular Momentum Transfer}
The evolution of finite disks  is qualitatively different to that of infinite disks, as evidenced by the decaying accretion
rate of Fig.~\ref{fig:torus_evolution}. Note, however, that while $\dott{M}_1$ and
 $\dott{M}_2$ also decay in time, they track the evolution $\dott{M}_{\rm b}$, with the
 ratio $\dott{M}_1/\dott{M}_2$ remaining roughly constant. Indeed, after normalizing by the running average
of $\dott{M}_{\rm  b}$  (top panel in Fig.~\ref{fig:finite_disk_accretion_1}),  the
 quantities $\langle \dott{M}_1\rangle_{30}/\langle \dott{M}_{\rm b}\rangle_{30}$
 and $\langle \dott{M}_2\rangle_{30}/\langle \dott{M}_{\rm b}\rangle_{30}$ remain roughly constant
 for $t\gtrsim100P_{\rm b }$. Furthermore, the angular momentum transfer {\it per unit accreted mass}
 $\langle \dot{J}_{\rm b}\rangle_{30}/\langle \dott{M}_{\rm b}\rangle_{30}$ 
 (middle panel in  Fig.~\ref{fig:finite_disk_accretion_1})  is also constant after the transient phase has ended.
 Similarly, the running averages of the migration rate per unit accreted mass 
  $\langle \dot{a}_{\rm b}\rangle_{30}/\langle \dott{M}_{\rm b}\rangle_{30}$ (bottom panel) rises  
  from a negative value at the beginning to a nearly constant value. 
   
   Remarkably, the values at which  $\langle \dot{J}_{\rm b}\rangle_{30}/\langle \dott{M}_{\rm b}\rangle_{30}$ 
   and   $\langle \dot{a}_{\rm b}\rangle_{30}/\langle \dott{M}_{\rm b}\rangle_{30}$  saturate are very
   close to the nominal values of $l_0$ and $\langle \dot{a}_{\rm b}\rangle/\dot{M_0}$ obtained
   from infinite disks (for $q_{\rm b}=0.5$,
   $l_0\approx0.75a_{\rm b}^2\Omega_{\rm b}$ and 
   $\langle \dot{a}_{\rm b}\rangle/\dott{M}_{0}\approx 2.4 a_{\rm b}/M_{\rm b}$; 
   see Fig.~\ref{fig:collected_results}). The same behavior is observed for other
    values of $q_{\rm b}$ (see Figs.~\ref{fig:finite_disk_accretion_2}-~\ref{fig:finite_disk_accretion_3}).
  Therefore, binaries accreting from finite mass reservoirs, while being  subject to decreasing accretion supplies,
  still accrete a consistent amount of angular momentum per unit accreted mass.

\section{The Role of Viscosity}\label{sec:viscosity}

In Sections~\ref{sec:infinite} and in our previous works \citep{mun16b,mun19},
we used a high viscosity coefficient $\alpha=0.1$ 
in order to the reach steady state of ``infinite'' disks for a small
number of binary orbits. The results of Section~\ref{sec:finite} suggest that
true steady state is not required to compute 
the angular momentum transfer rate per unit of accreted mass, and thus
high viscosities may not be necessary.
In this section, we provide an initial exploration of accretion
at lower viscosities.

\subsection{Low-viscosity Circumbinary Tori}
We setup a torus in the same way as described in Section~\ref{sec:finite}, imposing
a surface density profile as in Equation~(\ref{eq:initial_profile_finite}) but rescaling $\Sigma_0$ for each value
of $\alpha$ according to Equation~(\ref{eq:density_scaling}). We fix $q_{\rm b}=0.5$ for all simulations
and set $\alpha$ to be 0.01, 0.03 and 0.06. In Fig.~\ref{fig:viscosity_comparison}, we compare
the scaled surface density field at $t=0.5t_{\nu,{\rm cav}}$ for the four different viscosities explored
($t=100$, $160$, $320$ and $1000P_{\rm b}$ for
$\alpha=0.1$, 0.06, 0.03 and 0.01 respectively).
 In all cases,
  the outer structure of the CBD is essentially the same, as expected from viscous disk evolution. The binary cavity,
  on the other hand, appears to depend sensitively on the viscosity coefficient.   Low values
  of $\alpha$ also lead to larger cavities, as expected from torque balance arguments \citep{art94,mir15}. 
In addition, the lower the value of $\alpha$, the more
  complex the inflow structure, with multi-arm inflows being evident for $\alpha=0.01$  and $\alpha=0.03$
   \citep[see][]{far14,dor16,mos19}.  {A multi-ring structure in the CBD becomes increasingly prominent as viscosity is lowered. A similar behavior was noticed by \citet{dor13} (see their figure 10) and is attributed to the inability of local viscosity to damp away these disturbances. We have run an additional simulation with twice the number of cells and have obtained the same ringed structure.}

  The fact that the  bulk of the CBD evolves mostly in a viscous fashion implies that, to leading order, the mass transfer onto the binary
  depends on $\alpha$ only through a rescaled time coordinate 
  \begin{equation}\label{eq:dimensionless_time}
  {\cal T}=1+t/t_{\nu,{\rm cav}}
  \end{equation}
   \citep{lyn74,har98}. We show the evolution of
    $\langle\dot{J}_{\rm b}\rangle_{30}/\langle\dott{M}_{\rm b}\rangle_{30}$ for 
 $\alpha=0.06$,  $\alpha=0.03$  and $\alpha=0.03$ in Fig.~\ref{fig:viscosity_comparison_eigenvalue}, where the time coordinate
 is shown in conjunction with the dimensionless time ${\cal T}$. In these three cases, the evolution of the
 angular momentum transfer rate per unit mass is nearly homologous to the fiducial case with $\alpha=0.1$: there is a transient
 phase that lasts till ${\cal T}\approx1.5$, followed by a stationary phase in which   
 $\langle\dot{J}_{\rm b}\rangle_{30}/\langle\dott{M}_{\rm b}\rangle_{30}\sim$~constant. 
 For $\alpha\neq0.1$, 
 this constant is only slightly smaller than 
 the reference value of $l_0\approx0.75$ for an infinite disk (Fig.~\ref{fig:collected_results}), although in all cases,
 this asymptotic value is large enough to guarantee $\dot{a}_{\rm b}>0$ (cf. Equation~\ref{eq:migration_rate_ideal}).
 
{It is interesting to note that $\langle\dot{J}_{\rm b}\rangle_{30}/\langle\dott{M}_{\rm b}\rangle_{30}$ is
  barely changed as $\alpha$ is reduced.
The CBD and CSDs are responsible for negative and positive gravitational torques, respectively \citep{mun19}, and
  tidal truncation at low $\alpha$ suggests larger circumbinary cavities and smaller CSDs \citep{art94,mir15}. Perhaps surprisingly, these results suggest that, while the specific angular momentum transfer
  rate  may depend on $\alpha$, this dependence is weak}.


\section{Summary and Discussion}\label{sec:summary}
We have carried out 2D viscous hydrodynamical simulations of
circumbinary disk accretion using the finite-volume, moving-mesh code
{\footnotesize AREPO}. Focusing on circular binaries, we have considered various
binary mass ratios ($0.1 \leq q_{\rm b}\leq 1$), and studied accretion
from both ``infinite'' disks (with steady mass supply at large radii)
and finite-sized, viscously spreading disks (tori).  {\footnotesize AREPO}
allows us to follow the mass accretion through a wide radial extent of
the circumbinary disk, through accretion streams inside the binary
cavity and onto circum-single disks around the binary
components. This paper extends our recent studies of circumbinary
accretion (\citealp{mun16b,mun19}; see also \citealp{mir17}) which
focused on equal-mass (but generally eccentric) binaries accreting
from ``infinite'' disks. Our key findings are:

\begin{itemize}
\item[$(i)$] Binaries accreting from circumbinary disks generally gain
  angular momentum over long time scales
  (see Figs.~\ref{fig:collected_results}, \ref{fig:finite_disk_accretion_1}-\ref{fig:finite_disk_accretion_3} and~\ref{fig:viscosity_comparison_eigenvalue}). The quasi-steady accretion
  ``eigenvalue'' $l_0$, defined as the angular momentum transfer from
  the disk onto the binary per unit accreted mass, is robust against
  the radial extent of the surrounding disk/torus, lying in the range
  ($0.65$-$0.85)a_{\rm b}^2\Omega_{\rm b}$ for all mass ratios explored in
  this paper ($q_{\rm b}=0.1 - 1.0$) and depending weakly on the disk viscosity
  (for $\alpha$ between 0.01 and 0.1 and disk aspect ration $h_0=0.1$).
  {The corresponding migration rate $\langle\dot a_{\rm b}\rangle$  depends on $l_0$
 and the ``preferential accretion rate ratio'' $\eta=\langle\dott{M}_2\rangle/\langle\dott{M}_{\rm b}\rangle$. 
 For $q_{\rm b}\gtrsim 0.2$, we find that the binary expands in separation ($\langle\dot a_{\rm b}\rangle>0$).
For  $q_{\rm b} \lesssim 0.2$, $l_0 \approx l_{0,{\rm crit}}$, 
 which leads to a substantially reduced rate of outward migration, and possible inward migration.}
\item[$(ii)$] Starting from initial conditions with an empty binary cavity, all simulations
  exhibit an initial transient phase, after which the inner accretion flow
  and binary cavity settle into a quasi-steady state.
The duration and the inner flow structure 
of the transient phase are sensitive to the initial conditions and the fluid 
viscosity. This phase may be accompanied by angular momentum loss of the binary, 
but it does not represent the long-term behavior of binary-disk system.
\item[$(iii)$] 
{Even after the flow has reached the global quasi-steady state, the accretion 
onto the central binary components is highly variable on short time scales. 
The dominant variability frequency changes from 
$0.2\Omega_{\rm b}$ at $q_{\rm b}>0.5$ to $\Omega_{\rm b}$ 
at $q_{\rm b}<0.5$ (see Figures~\ref{fig:collected_frequencies}-\ref{fig:collected_results}).}
\item[$(iv)$] The low-mass component ($M_2$) of the binary generally
  accretes more mass (see Figs.~\ref{fig:collected_results}, \ref{fig:finite_disk_accretion_1}
  and \ref{fig:finite_disk_accretion_2}). In the quasi-steady state, the
  time-averaged accretion fraction $\langle\dott  M_2\rangle/\langle
  \dott  M_b\rangle$ increases from $50\%$ at $q_{\rm b}=1$ to $90\%$ at
  $q_{\rm b}=0.1$.

\end{itemize}

\subsection{Comparison to Previous Works}
The findings presented in this paper and in \citet{mun19},  
(using {\footnotesize AREPO}), as well as in \citet{mir17}  (using {\footnotesize PLUTO})
and \citet{moo19} (using {\footnotesize ATHENA++}), contradict the long-standing notion
that binaries lose angular momentum to circumibinary disks and shrink its orbit.
{ These works combined contradict the claim by \citet{tan17} that only
an ``unphysically fast'' mass extraction rate by the ``sinks'' (binary components) can produce positive net torques.}
In \citet{mun19} (see their Section 5.2.1), we have provided a detailed
comparison to previous works, and discussed the possible reasons why
some other works have led to different results. 
  Overall, ignoring
issues associated with calculation and numerical methods,
there are three possible causes: (i) Some earlier works adopted the erroneous
assumption that the binary suppresses mass accretion due to tidal
torques and that only the gravitational torque from the CBD determines
the evolution of the binary; (ii) Some simulation studies were of
short-term duration (in relation to the viscous time), making them
representative of the transient phase; (iii) Some studies
considered very massive CBDs, for which the binary may evolve quickly
before the disk has time to relax to a quasi-steady state.

Regarding point (ii) above: our simulations of accretion from finite disks (tori)
(see Section~\ref{sec:finite}) clearly identified a transient phase where the binary
loses angular momentum and undergoes inward migration, followed by a
quasi-steady phase with outward migration. This
transient phase lasts longer for low-viscosity disks.  This explains
why some low-viscosity simulations appear to induce binary shrinkage,
simply because the CBD has not had time to fill in the cavity \citep[e.g.][]{rag16}.
 Regarding (iii): Our simulations (as well as those in \citealp{mun16b},
\citealp{mir17} and \citealp{moo19}) considered
non-self-gravitating disks, and required that the local disk mass (at
a few times $a_{\rm b}$) be much less than the binary mass.  There may be
situations where this condition is not satisfied.  The massive CBDs
with empty initial cavities explored by \citet{cua09} (in the context
of MBBHs following galaxy mergers)
could be an example: the binary runs away from the initial
configuration before the CBD can fill the cavity \citep[see also][]{esc05,roe11}.

{While it is known that 3D effects will alter the excitations of waves in the CBD and CSDs
via Lindblad resonances \citep[at least for small $q_{\rm b}, $e.g.,][]{tan02,bat03}},
it is unlikely that the 2D nature of our simulations plays an important role in the sign
of $\langle{\dot a}_{\rm b}\rangle$. \citet{moo19} have 
already shown that, for a 3D CBD
coplanar with the binary,  the measured value of $l_0$ is 
consistent with the 2D results of \citet{mun19}; for misaligned CBDs, their
 eigenvalue is significantly larger, i.e., outward migration is even faster.
{We acknowledge, nevertheless, that  \citet{moo19} focused on the $q_{\rm b}=1$ case; thus, 
the qualitative consistency between 2D and 3D remains untested for $q_{\rm b}\neq 1$.}
The equation of state (EoS) assumed could have a moderate effect on our results by modifying
the angular momentum currents \citep{lee16,mir19}, 
although it is unclear whether
a more realistic EoS \citep[see, e.g.,][]{kle19} would make $l_0$ larger or smaller.
Additional physics, such as the launching of a magnetized winds \citep[e.g.,][]{bla82},
could modify the angular momentum transfer rate. While an intriguing possibility, winds also depress the net
  accretion rate by carrying away mass, and their net effect on the net value 
  of $\langle\dot{J}_{\rm b}\rangle/\langle\dott{M}_{\rm b}\rangle$ is difficult to anticipate.

Finally, we note that, during the revision stage of this manuscript, a preprint by
\citet{duf19} was posted online, carrying out a similar parameter-space exploration as 
this work and finding qualitatively similar results as ours.

\subsection{Implications for Supermassive Binary Black Hole Coalescence}
Our finding that binaries gain angular momentum from circumbinary accretion
casts doubt on the commonly accepted role of gas disks in
MBBH orbital evolution and coalescence \citep[see][for comprehensive 
reviews on the topic]{dot12,col14}. 
It also impacts the gravitational wave (GW) emission from possible
mergers of MBBHs in galactic nuclei.
Such mergers produce GWs anywhere from the 
LISA band ($f=\Omega_{\rm b}/\pi\sim10^{-4}$~Hz for $M_{\rm
  b}\lesssim10^7M_\odot$) to the Pulsar Timing Array (PTA) and SKA
bands ($f\sim10^{-7}$~Hz for $M_{\rm b}\gtrsim10^9M_\odot$).  
Since binaries like the ones explored in this work
( $q_{\rm b}\gtrsim0.4$) dominate the GW background at
frequencies $\lesssim10^{-7}$~Hz \citep{kel17b}, if such 
MBBH merging events are suppressed by orbital stalling, the expected GW
background would be significantly altered. 
It has been argued that loss-cone filling in the most massive galactic merger 
remnants is efficient enough to bring MBBHs into the GW regime
\citep{kha11}, or that multiple mergers might assist binary coalescence via triple
 interactions \citep[e.g.][]{hof07}. 
 A quantitative study of all competing effects acting
simultaneously on the MBBHs would be useful.

If comparable-mass MBBHs are stalling or expanding due to CBD
torques, we may expect a (as yet undetected) population of binary/dual
black holes in galaxies that have undergone major mergers.  Possible
MBBH candidates include the AGN in 0402+379 \citep{rod06} and the
quasar OJ 287 \citep{val08}.  Observational campaigns to search for
signatures of binarity (or ``duality'') using photometric
variabilities \citep[e.g.][]{gra15a,gra15b,liu16,cha16} may reveal
more MBBH candidates.

\subsection{Implications for Stellar Binary Formation}
In contrast to MBBHs, stellar binaries are likely to have formed
 within a massive disk, in a process of gravitational fragmentation followed
  by migration  \citep[e.g.,][]{bon94b,bat97,kra08}.
  The positive migration rates found for $q_{\rm b} \gtrsim 0.3$
   represent an obstacle to this qualitative picture of binary formation. On the other hand, outward migration might help explain why binaries stop hardening before complete coalescence.
    The drastic reduction in the magnitude of $\langle\dot a_{\rm b}\rangle$
     found at $q_{\rm b} \approx 0.2$ (Figure~\ref{fig:collected_results})
      may offer a solution to this conundrum. A transition in the sign of of
$\langle\dot a_{\rm b}\rangle$ is to be expected at small enough mass ratios, as it is known from linear theory that migration is inward for $q_{\rm b} \ll 1$ and no secondary accretion \citep[e.g.,][]{war86,war97}.

We underscore that $l_0 > 0$ is not synonymous with outward migration: the fact that
$\langle\dot a_{\rm b}\rangle>  0$ in Figure~\ref{fig:collected_results} 
results from $l_0 >  l_{0,{\rm crit}}$ (Equation~\ref{eq:migration_rate_ideal}). 
Since $l_0$ is roughly constant for $0.1 < q_{\rm b} < 1$, and $l_{0,{\rm crit}}$ increases for smaller 
$q_{\rm b}$ (due to increasing $\eta$), inward migration may be possible for $q_{\rm b} \lesssim 0.2$.
An additional transition from $l_0 > 0$ to $l_0 < 0$ may occur at an even smaller mass ratio. 
Indeed, for planetary mass companions ($q_{\rm b} \ll 1$, $\eta \approx 0$), 
\citet{dem19}
reported negative eigenvalues $l_0 \approx -4 ({q_{\rm
    b}}/{\alpha }) \left({h_0}/{0.05}\right)^{-2}a_{\rm
  b}^2\Omega_{\rm b}$ for massive planets ($2\times10^{-3}\gtrsim
q_{\rm b}\gtrsim10^{-4}$) embedded in low-mass
 steady-state disks, which evolve in a ``modified'' Type-II migration scenario
(\citep{dur15};  although see \citealp{duf14}).

{Alternatively, a plausible solution to the positive torque conundrum is that the binary separation is largely set
early on during the Class 0 phase of star formation \citep{bat00}.
Quasi-spherical accretion from an envelope of low specific angular momentum will naturally
allow for large values of $\langle\dot{M}_{\rm b}\rangle$ while keeping $\langle\dot{J}_{\rm b}\rangle$ low,
resulting in $\langle \dot{a}_{\rm b}\rangle\lesssim0$ (Equation~\ref{eq:migration_rate_ideal}). Our simulations
would represent a later, disk-dominated viscously evolving phase. The transition from envelope-dominated
accretion to disk-dominated accretion might result in a sign reversal in $\langle\dot{a}_{\rm b}\rangle$ 
\citep[see fig. 3 in][]{bat00}.}


\acknowledgments{
The authors thank Volker Springel for making the {\footnotesize AREPO} code
available for this work.
DJM thanks Enrico Raguzza, Matthew Bate and Luke Z. Kelley for illuminating conversations and
Adam Dempsey and Yoram Lithwick for many valuable technical discussions.

This work has been supported in part by NASA grants NNX14AG94G and 80NSSC19K0444,
 NSF grant AST-1715246 (Cornell), and NASA grant ATP-170070 (Arizona).
D.J.M. acknowledges support by the computational resources and staff contributions provided for the
 Quest high performance computing facility at Northwestern University which is jointly supported by the Office of the Provost, 
 the Office for Research, and Northwestern University Information Technology.}

\bibliographystyle{apj}

\end{document}